\documentclass[twocolumn,tighten]{aastex63}
\usepackage{graphicx}
\usepackage{amsmath}
\usepackage{natbib}
\usepackage{amssymb}

\usepackage{sidecap}

\def\KB{k_{\rm B}}
\def\MP{m_{\rm p}}
\def\Msun{{\mathrm{M}_\odot}}
\def\Rsun{{\mathrm{R}_\odot}}

\def\Msun{\ensuremath{{M}_\odot}}
\def\Rsun{\ensuremath{{R}_\odot}}
\def\lumi{\ensuremath{\mathrm{erg\,s^{-1}}}}

\def\kms{\ensuremath{\mathrm{km\,s^{-1}}}}
\def\smyr{\ensuremath{M_{\odot}\mathrm{\,yr^{-1}}}}
\def\dR{\ensuremath{\Delta R}}
\def\dt{\ensuremath{\Delta t}}

\shorttitle{Light Curve Model for Supernova Precursors}
\shortauthors{Matsumoto \& Metzger}

\begin{document}

\newcommand{\be}{\begin{equation}}
\newcommand{\ee}{\end{equation}}

\title{Supernova Precursor Emission and the Origin of Pre-Explosion Stellar Mass-Loss }

\author{Tatsuya Matsumoto}
\affil{Department of Physics and Columbia Astrophysics Laboratory, Columbia University, Pupin Hall, New York, NY 10027, USA}

\author[0000-0002-4670-7509]{Brian D. Metzger}
\affil{Department of Physics and Columbia Astrophysics Laboratory, Columbia University, Pupin Hall, New York, NY 10027, USA}
\affil{Center for Computational Astrophysics, Flatiron Institute, 162 5th Ave, New York, NY 10010, USA}

\begin{abstract}
A growing number of core collapse supernovae (SNe) which show evidence for interaction with dense circumstellar material (CSM) are accompanied by ``precursor'' optical emission rising weeks to months prior to the explosion.  The precursor luminosities greatly exceed the Eddington limit of the progenitor star, implying they are accompanied by substantial mass-loss.  Here, we present a semi-analytic model for SN precursor light curves which we apply to constrain the properties and mechanisms of the pre-explosion mass-loss.  We explore two limiting mass-loss scenarios: (1) an ``eruption'' arising from shock break-out following impulsive energy deposition below the stellar surface; (2) a steady ``wind'' due to sustained heating of the progenitor envelope. The eruption model, which resembles a scaled-down version of Type IIP SNe, can explain the luminosities and timescales of well-sampled precursors, for ejecta masses $\sim 0.1-1\,\Msun$ and velocities $\sim 100-1000\,\rm km\,s^{-1}$. By contrast, the steady-wind scenario cannot explain the highest precursor luminosities $\gtrsim10^{41}\,\rm erg\,s^{-1}$, under the constraint that the total ejecta mass not exceed the entire progenitor mass (though the less-luminous SN 2020tlf precursor can be explained by a mass-loss rate $\sim1\,\Msun\,\rm yr^{-1}$). However, shock interaction between the wind and pre-existing (earlier ejected) CSM may boost its radiative efficiency and mitigate this constraint.  In both eruption and wind scenarios the precursor ejecta forms compact ($\lesssim10^{15}$ cm) optically-thick CSM at the time of core collapse; though only directly observable via rapid post-explosion spectroscopy ($\lesssim$ few days before being overtaken by the SN ejecta), this material can boost the SN luminosity via shock interaction.
\end{abstract}

\section{Introduction}
\label{sec:introduction}

A fraction of massive stars undergo strongly enhanced mass-loss near the very ends of their lives, forming a dense circumstellar medium (CSM) around themselves (e.g., \citealt{Smith2014}).  When the stars explode as supernovae (SNe), the CSM manifests through narrow emission lines in the optical spectra generated by the as-yet-unshocked slowly expanding CSM, photoionized by the SN light (e.g., SN Type IIn, Ibn or Icn, depending on whether the spectrum is hydrogen-rich, hydrogen-poor but helium-rich, or both hydrogen- and helium-poor, respectively; \citealt{Schlegel90,Filippenko1997,Foley+07,Pastorello+08,Nyholm+20,Gal-Yam+22,Fraser+21}).
The related technique of ``flash spectroscopy'' (e.g., \citealt{GalYam+2014,Khazov+2016}) demonstrates that the progenitor's mass-loss rate is elevated just before core collapse, among a significant fraction of even nominally CSM-free SNe (e.g., \citealt{Yaron+2017,Bruch+2021}).

The light curves of some of the most luminous SNe are powered at least in part by shock interaction between the SN ejecta and dense CSM released from the progenitor star in the days to weeks to years prior to its terminal core collapse (e.g., \citealt{Smith&McCray07,Chevalier&Irwin11,Ginzburg&Balberg12,Svirski+12,McDowell+18,Suzuki+19}).  At typically larger radii and lower densities, CSM shock interaction can also produce thermal and non-thermal X-ray and radio emission (e.g., \citealt{Chandra+12,Margutti+14,Dwarkadas+16,Chakraborti+16,Margutti+17,Chiba+20}).

If the pre-SN stellar mass-loss is modeled as a steady wind, the inferred mass-loss rates in Type IIn SNe,  $\dot{M}\sim10^{-4}-0.1\smyr$ \citep{Fox+2011,Kiewe+2012,Moriya+2014}, are significantly larger than can be explained by line-driven winds $\dot{M}\lesssim10^{-5}\smyr$ \citep{Vink+01,Smith2014}. Instead, several alternative physical processes have been proposed to give rise to enhanced mass-loss just prior to core collapse.  Mass-loss can occur due to intense heating of the stellar envelope by damping of waves excited by vigorous convection in the stellar core (e.g., \citealt{Quataert&Shiode2012,Shiode&Quataert2014,Fuller2017,Fuller&Ro2018,Leung&Fuller2020,Wu&Fuller22}).  In this scenario, the timing of the mass-loss should correspond to late stages of nuclear burning, which occurs on timescales of days to weeks prior to explosion for silicon burning; to several years for oxygen and neon burning (e.g., \citealt{Woosley+2002}).  Another mechanism for generating pre-SN mass-loss invokes sudden energy release deep inside the star due to instabilities associated with late-stages of nuclear shell burning (e.g., \citealt{Meakin&Arnett07,Smith&Arnett14,Fields&Couch21,Varma&Muller21,Yoshida+21}).  Mass-loss due to interaction with a close binary companion may also play a role in some events (e.g., \citealt{Chevalier12,Mcley&Soker2014,Sun+20}). In very massive rotating stars which experience efficient mixing, centrifugally-induced mass-loss may accompany contraction during the star's final burning stages (e.g., \citealt{Aguilera-Dena+18}).

``Precursor'' outbursts prior to the main SN explosion (e.g., \citealt{Ofek+2014c}) offer a clue to distinguishing these various mass-loss mechanisms.  Several precursor events accompanying interacting SNe have been observed in recent years typically weeks to months prior to the main explosion, the most well-studied of which accompanied SN 2009ip \citep{Fraser+2013,Mauerhan+2013,Pastorello+13,Prieto+2013,Graham+2014,Margutti+14,Mauerhan+2014,Levesque+2014,Martin+2015,Graham+2017,Reilly+2017,Smith+22}, 2010mc \citep{Ofek+2013b}, 2015bh \citep{EliasRosa+16,Ofek+2016,Thone+2017,Jencson+2022}, 2016bhu \citep{Pastorello+18}, LSQ13zm \citep{Tartaglia+16}, and 2020tlf \citep{JacobsonGalan+2022} as well as the implication of pre-SN activity in SN 2010bt \citep{Elias-Rosa+18}, 2013gc \citep{Reguitti+19}, 2018cnf \citep{Pastorello+19b}, 2019zrk \citep{Fransson+2022}, 2021foa \citep{Reguitti+2022}, and 19 events compiled in \cite{Strotjohann+21}.  Efforts to monitor a large number of massive stars in nearby galaxies to determine which explode as SNe \citep{Kochanek08} and systematic analyses of interacting SNe \citep{Ofek+2014c,Bilinski+2015,Strotjohann+21}, rule out luminous precursors accompanying all SNe \citep{Kochanek+2017,Johnson+2018}. However, \cite{Strotjohann+21} found that $\simeq25\,\%$ of Type IIn SNe (themselves accounting for $\simeq 10\,\%$ of core collapse SNe; e.g., \citealt{Perley+20}) exhibit precursors brighter than $-13$ mag (or luminosity $\gtrsim 5\times10^{40}\,\lumi$) three months before the SN explosion.

\begin{figure}
\begin{center}
\includegraphics[width=85mm, angle=0,bb=0 0 285 214]{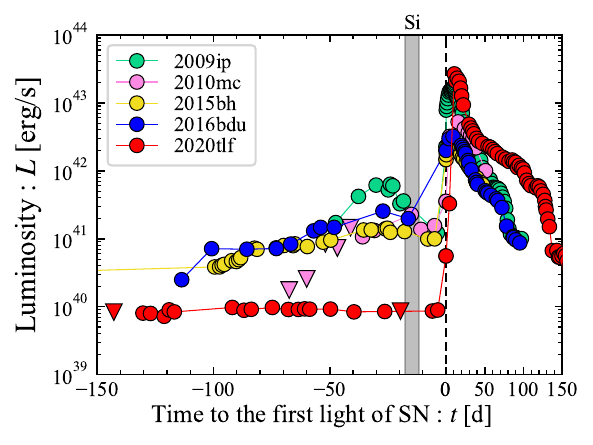}
\caption{Sample of SN precursor light curves, showing the luminosity as a function of time relative to the onset of the SN (which we define by the luminosity rising abruptly by a factor of $\gtrsim 10$).  Shaded region shows the characteristic timescale range before core collapse corresponding to core silicon burning in $13-20\,\Msun$ stars \citep{Woosley+2002}. References to the light curve data are given in Table~\ref{table:obs}.}
\label{fig:obs}
\end{center}
\end{figure}

Figure~\ref{fig:obs} and Table~\ref{table:obs} summarize the bolometric light curves and other observable properties of a set of well-sampled SN precursors.  Most of these SNe are classified as Type II, so we focus in this work on precursors from hydrogen-rich stars (though precursor outbursts have also been observed in some hydrogen-poor SNe; e.g., \citealt{Foley+07,Pastorello+08}).\footnote{Some massive stars also generate bright eruptions$-$so-called ``giant eruptions'' in luminous blue variable phase or ``SN imposters'' (e.g., \citealt{Pastorello+10,Pastorello&Fraser2019}), which are potentially related to instabilities in massive star envelopes (e.g., \citealt{Humphreys&Davidson94,Owocki15,Jiang+18}) and accompanied by mass-loss \citep[e.g.,][]{Gal-Yam+07,Gal-Yam&Leonard2009}, but typically are not coincident with the terminal core collapse event.  Here we focus on SN precursors which precede the terminal explosion by at most a few years and thus are likely to be causally connected to late-stages of nuclear burning.}  The luminosities of the observed precursors $\sim 10^{40}-10^{41}$ erg s$^{-1}$ are typically well above the Eddington limit for typical SN progenitor stars of mass $\simeq 10-20\,\Msun$,\footnote{The progenitors of SN 2009ip and 2015bh may be more massive $\gtrsim35-50\,\Msun$ \citep{Smith+2010,Foley+2011,Boian&Groh2018}.} indicating that this emission phase is accompanied by substantial mass-loss, which should then contribute to CSM interaction with the supernova light or ejecta in the post-explosion phase.

In this paper, we present a model for the light curves of SN precursors from hydrogen-rich stars, which we apply to the observed precursor sample in order to constrain the pre-SN mass-loss phase.  With the exception of radiation hydrodynamical simulations by \citet{Dessart+2010}, few previous theoretical works have attempted to model the precursor emission phase.  Instead, most constraints on the mass and radial distribution of the CSM around SN progenitors have come from the post-explosion interaction (e.g., \citealt{Smith&McCray07,Chevalier&Irwin11,Chatzopoulos+13,McDowell+18,Suzuki+19}).  Though our light curve model makes several simplifying assumptions, these same features make it flexible and applicable to a wide parameter space of potential pre-explosion mass-loss behavior.  As we shall discuss, we find that the bulk of the CSM generated during the observed precursor outbursts will usually be located so close to the progenitor star at the time of the explosion as to be quickly overtaken by the SN ejecta, making this material challenging to directly observe by other means.  

This paper is organized as follows.  In Section \ref{sec:model} we introduce two idealized models for the nature of the pre-SN mass-loss phase (dynamical ``eruption'' versus steady ``wind'') and describe our model for calculating the precursor light curve from each case.  In Section \ref{sec:result} we present the results of our light curve calculations and interpret them analytically.  We also assess the ability of the eruption and wind models to explain the observed SN precursor events and determine the ejecta parameters required in each case. We discuss the resulting CSM from precursor events in Section \ref{sec:discussion} and finally  summarize our results and conclude in Section \ref{sec:summary}.

\begin{table*}
\begin{center}
\caption{Observed Properties of SN Precursor Emission}
\label{table:obs}
\begin{tabular}{lccccc}
\hline
Event&$L_{\rm pre}^{(a)}$&$t_{\rm pre}^{(b)}$&$E_{\rm pre}^{(c)}$&$v_{\rm obs}$&Ref.\\
&[erg/s]&[d]&[erg]&[$10^3$ km/s]\\
\hline
SN 2009ip&$3.6\times10^{41}$&44$^\P$&$1.4\times10^{48}$$^\P$&0.8-1.4, 8-9, 14-15$^\ddagger$&1,2,3\\
SN 2010mc&$1.6\times10^{41}$&31&$4.4\times10^{47}$&1-3 (6 d)$^\dagger$&4\\
SN 2015bh&$9.3\times10^{40}$&95$^\P$&$7.6\times10^{47}$$^\P$&0.6-1, 0.9-1.5, 2.6-6$^\ddagger$&5\\
SN 2016bdu&$1.2\times10^{41}$&97$^\P$&$1.0\times10^{48}$$^\P$&0.4 ($\simeq10$ d)$^\dagger$&6\\
SN 2020tlf&$8.8\times10^{39}$&127&$9.7\times10^{46}$&0.05-0.2 (10 d)$^\dagger$&7\\
\hline
\multicolumn{6}{l}{$^{(a)}$ Average luminosity of precursor emission, defined as $L_{\rm pre} = E_{\rm pre}/t_{\rm pre}$.}\\
\multicolumn{6}{l}{$^{(b)}$ Duration from the first precursor detection to the time of the SN explosion.}\\
\multicolumn{6}{l}{$^{(c)}$ Total radiated energy of the precursor emission.}\\
\multicolumn{6}{l}{$^\P$Lower limit due to the lack of detection or flux upper limit.}\\
\multicolumn{6}{l}{$^\ddagger$Obtained via spectroscopy during the precursor emission itself. The H$\alpha$ line}\\
\multicolumn{6}{l}{profile is fitted by multiple velocity components.}\\
\multicolumn{6}{l}{$^\dagger$Obtained via flash-spectroscopy $t_{\rm flash}$ (shown in parentheses) after the SN}\\
\multicolumn{6}{l}{explosion. Note that these velocities are sometimes too low to be associated with}\\
\multicolumn{6}{l}{mass ejection responsible for the precursor emission because the latter would have}\\
\multicolumn{6}{l}{been overtaken by the SN ejecta by the time of the flash-spectroscopy ($t_{\rm flash}$).}\\
\multicolumn{6}{l}{{\bf Ref.} 1: \citet{Mauerhan+2013}, 2: \citet{Pastorello+13},}\\
\multicolumn{6}{l}{3: \citet{Margutti+14}, 4: \citet{Ofek+2013b}, 5: \citet{EliasRosa+16},}\\
\multicolumn{6}{l}{6: \citet{Pastorello+18}, 7: \citet{JacobsonGalan+2022}.}\\
\end{tabular}
\end{center}
\end{table*}

\section{Precursor Emission Model}\label{sec:model}

We calculate SN precursor light curves largely following the semi-analytical model developed in \citet{Matsumoto&Metzger2022,Metzger+2021}.  The general approach is as follows: the pre-SN mass-loss is divided into multiple ejecta shells ordered based on their velocity, and for each shell we calculate its thermal evolution and resulting emission by means of a one-zone model.  The total light curve is then obtained by summing the luminosity contributions from each shell.  For simplicity we assume the ejecta to be spherically symmetric, even though in some mass-loss scenarios it could possess a non-spherical (e.g.,  equatorial disk-like) geometry.

A key input to the model is the velocity distribution of the pre-SN ejecta.  We explore two physically-motivated scenarios for its form, which bracket extremes along a continuum of possible behaviors.
\begin{enumerate}
    \item An ``eruption,'' in which mass-loss occurs in a single event on a timescale comparable or less than the dynamical time at the ejection radius.  Physically, a temporally-concentrated injection of energy deep inside the star generates a shock wave, which propagates radially outwards and accelerates the stellar envelope, unbinding a portion of its mass \citep{Dessart+2010,Kuriyama&Shigeyama2020,Linial+2021,Ko+2021}.  Although we remain indifferent to the origin of the sudden energy injection, one possible physical realization would be a dynamical instability associated with unstable nuclear shell burning.
    \item A continuous ``wind,'' in which mass-loss occurs from the star at a roughly constant rate, over timescales much longer than the dynamical time.  Physically, this is expected to occur when the stellar envelope is heated well above the Eddington luminosity at a roughly constant rate (e.g., \citealt{Smith&Owocki06,Quataert+2016}). This scenario may be approximately realized in wave-heating scenarios (e.g., \citealt{Quataert&Shiode2012,Fuller2017}).
\end{enumerate}

The following subsections describe the eruption and wind scenarios individually.  Throughout this work we assume that the precursor emission arises directly from a single mass-loss event or mass-loss phase prior to the terminal explosion; however, we caution that some events may arise from more complicated circumstances (e.g., multiple episodes of mass-loss giving rise to collisions between ejecta shells; dim SNe followed by bright CSM interaction; or binary star mergers) in particular for SN 2009ip (e.g., \citealt{Pastorello+13,Soker&Kashi2013,Smith+2014}).

\subsection{Eruption Scenario}
\label{sec:eruption_setup}

In the eruption scenario, energy is injected suddenly into the stellar envelope, driving a shock wave towards its surface (e.g., \citealt{Dessart+2010}).  As a result, we assume the ejecta achieve a homologous density profile $\rho(v = r/t)$, defined such that $M_{\rm ej}(>v) = 4\pi \int_{v}^{\infty}\rho(v)r^{2}dr$ is the ejecta mass above a given velocity $v$.

Depending on the magnitude of the injected energy, the ejecta can be accelerated to an arbitrarily high velocity as long as the latter exceeds the escape speed at the radius $R_0$ matter is ejected (e.g., \citealt{Linial+2021}),
\begin{align}
v_{\rm esc}&=\sqrt{\frac{2GM_\star}{R_0}}\\
&\simeq200{\,\rm km\,s^{-1}\,}\biggl(\frac{M_\star}{10\,\Msun}\biggl)^{1/2}\biggl(\frac{R_0}{10^{2}\,\Rsun}\biggl)^{-1/2}
	\label{eq:v_esc}\ ,
\end{align}
where $G$ is the gravitational constant and $M_\star$ is the progenitor mass.  Spectroscopic observations of SN precursors, in the few cases available, indicate high ejecta speeds $\sim 10^{2}-10^{3}$ km s$^{-1}$ (Table~\ref{table:obs}).  This suggests the deposited energy being comparable to the stellar binding energy if released at radii $R_0 \sim 1-100 \,\Rsun$ deep below the surface $R_{\star} \sim 10^{3}\,\Rsun$ of a RSG progenitor (or with energy greatly exceeding the local binding energy if deposited closer to the surface).

We now describe our method to calculate the emission from a single shell of mass $M$ and velocity $v$ ejected at $t=0$ from radius $R_0$.
The radial distance of the shell from the progenitor's center is given by $R=v\cdot t+R_0$ and its volume is given by $V=4\pi R^2\dR$, where $\dR$ is the width of the shell.
For a homologously expanding ejecta, the width is given by $\dR=dv \cdot t+\dR_0$, where $dv$ is the velocity difference from the next shell and $\dR_0$ is the initial width of the shell.  The initial internal energy of the shell is assumed to equal its kinetic energy $E_0=Mv^2/2$.\footnote{In detail, for this initial condition, the shell will accelerate after the ejection by $PdV$ work within several dynamical times $\sim R_0/v$; however, since the increase of the velocity is by at most a factor of $\sqrt{2} \simeq1.4$, we neglect this acceleration phase and assume the shells expand at constant velocity from $t = 0$.}  The internal energy is comprised of ideal gas, radiation, and ionization energy:
\begin{align}
E&=\frac{3}{2}(1+\bar{x})N\KB T+aT^4V+\sum_{i}NA_ix_i\varepsilon_{i}\ ,
	\label{eq:internal_energy}
\end{align}
where $\bar{x}=\sum_i A_ix_i$, $A_i$, $x_i$, and $\varepsilon_i$ are the mean ionization degree, abundance fraction, degree of ionization, and ionization energy of species $i$, respectively.  We calculate $\bar{x}$ and $x_{i}$ by solving the Saha equation taking into account singly-ionized hydrogen and helium. The other quantities are the total number of nuclei $N$, the Boltzmann constant $\KB$, temperature $T$, and radiation constant $a$.

The thermal evolution of the ejecta is described by the first law of thermodynamics:
\begin{align}
\frac{dE}{dt}=-(\gamma_3-1)\frac{E}{V}\frac{dV}{dt}-L\ ,
	\label{eq:thermo}
\end{align}
where $\gamma_3$ and $L$ are the adiabatic index and radiated luminosity, respectively.
In the adiabatic loss term, we have used the fact the pressure is given by $P=(\gamma_3-1)E/V$.
The adiabatic index is calculated from the density, temperature, and the ionization state of the shell as in \citet{Matsumoto&Metzger2022,Kasen&RamirezRuiz2010}.  The radiative loss term is approximated by the photon diffusion luminosity
\begin{align}
L=\frac{E_{\rm rad}}{S(t)t_{\rm d}+t_{\rm lc}}\ ,
	\label{eq:luminosity}
\end{align}
where $E_{\rm rad}=aT^4V$ is the radiation's internal energy, $t_{\rm d}$ is the photon diffusion time, and $S(t)$ is a suppression factor discussed below.  The diffusion time is that over which photons escape radially through the ejecta,
\begin{align}
t_{\rm d} = \frac{R\tau}{c}\,\,\,\text{and}\,\,\, \tau = \int_R \kappa \rho dr\ ,
	\label{eq:t_diff}
\end{align}
where $c$ is the speed of light and $\kappa(\rho,T)$ is Rossland-mean opacity, which we approximate using the analytic expression provided in \cite{Matsumoto&Metzger2022} for solar metallicity composition material.  The dominant form of opacity near peak light is electron scattering  (under conditions of partial or full ionization), though Kramer's opacity can become relevant at high densities and low temperatures.  The suppression factor,  
\begin{align}
S(t) \equiv\frac{e^{t_{\rm d}/t}-1}{t_{\rm d}/t}\ ,
    \label{eq:suppression}
\end{align}
acts to reduce the radiative losses exponentially for $t\ll t_{\rm d}$.  Without this correction, the luminosities of all radiation-pressure-dominated shells reach a value $L\propto R_0 v/\kappa$ starting immediately after their ejection and contribute almost equally (up to a factor of $v$) regardless of their optical depth; this behavior is unphysical, however, because at $t\lesssim t_{\rm d}$ most photons are still trapped within ejecta and only the tiny fraction in the ``diffusion tail'' can escape \citep[see also][]{Piro&Nakar2013}. The light crossing time $t_{\rm lc}=R/c$ in the denominator of Eq.~\eqref{eq:luminosity} limits the photon escape timescale from the system at late times.

The mass profile of the ejecta depends on the details of the ejection process. The process of shock-breakout motivates a power-law profile  \citep{Nakar&Sari2010},
\begin{align}
M_{\rm ej}(>v)&=M_{\rm ej}(v/v_{\rm ej})^{-\beta}\ ,\,\, v > v_{\rm ej}\ ,
	\label{eq:mass_profile}
\end{align}
where $M_{\rm ej}$ is the total ejecta mass and $v_{\rm ej}$ is the minimum velocity.  
Assuming the stellar envelope can be described as a polytrope of index $n$, and that the energy injection occurs close to the stellar surface, then we can set the initial radius $R_0\sim R_\star$ for all shells and the process of shock breakout will impart a self-similar velocity profile of the form $v\propto \rho^{-\mu}$ \citep{GandelMan&FrankKamenetskii1956,Sakurai1960}, where $\mu\simeq0.22$ (for $n=3/2$) and $\mu=0.19$ (for $n=3$).  Since the density profile and external mass in this scenario obey $\rho \propto x^{n}$ and $M(>r)\propto \rho x R_\star^2$, where $x=R_\star-r$ is the depth measured from the surface, the power-law index entering Eq.~(\ref{eq:mass_profile}) in this scenario becomes $\beta=(n+1)/{\mu n}\simeq7.6$ for $n=3/2$ and 7.0 for $n=3$. 
The shock breakout solution also gives a relation between the velocity and initial depth of each shell $v=v_{\rm ej}(x/x_{\rm ej})^{-\mu n}$, where the depth of the innermost slowest shell ($v=v_{\rm ej}$) obeys $x_{\rm ej}/R_\star\sim (M_{\rm ej}/M_\star)^{\frac{1}{n+1}}$.
The initial width of each shell is likewise given by 
\begin{align}
\dR_0(v)=\biggl|\frac{dx}{dv}\biggl|dv=\frac{x_{\rm ej}}{\mu n v_{\rm ej}}\biggl(\frac{v}{v_{\rm ej}}\biggl)^{-\frac{\mu n+1}{\mu n}}dv\ .
	\label{eq:shell width}
\end{align}
Note that above prescription holds only for energy injected into layers of the star close to the surface ($R_0\sim R_\star$ and $M_{\rm ej}\ll M_\star$). Nevertheless, in what follows we shall apply $R_0=R_\star$ also to the eruptions with greater ejecta masses, seeded by energy deposition at deeper layer $\ll R_\star$. Realistically, after the shock passage, each mass shell is quickly imparted comparable internal and kinetic energies, and hence the initial radius $R_0$ should be smaller than $R_\star$ and different for each shell. While this complicated hydrodynamics should be studied by numerical approaches with more specific stellar models to set more accurate initial conditions, we assume for simplicity that all shells are ejected at $t=0$ from $R_\star$ with equal internal and kinetic energies as adopted for analytical light curve modeling of SNe \citep{Arnett1980,Arnett1982,Popov1993}. Physically, heating near the surface will occur as the ejecta from deeper layers collides and shocks with material closer to the surface.

In summary, the eruption scenario is described by four main parameters: total ejecta mass $M_{\rm ej}$, initial radius $R_0$ (which we canonically set to $R_0=R_\star$, motivated by the shock-breakout prescription), minimum ejecta speed $v_{\rm ej}$ (which must exceed $v_{\rm esc}$; Eq.~\ref{eq:v_esc}), and the time of the eruption prior to the SN explosion, $t_{\rm erupt}$.  The power-law index of mass profile $\beta$ is also a free parameter, though we shall take $\beta=7.6$ as fiducial (corresponding to RSG progenitors), motivated by above discussion.  Although the self-similar shock breakout solution described above holds only near the stellar surface, we find that the light curve properties are not sensitive to the precise value of $\beta$ for otherwise fixed values of $M_{\rm ej}$ and $v_{\rm ej}$ (see \citealt{Matsumoto&Metzger2022}; their Fig.~6).  We assume initial shell widths given by Eq.~\eqref{eq:shell width}; however, the calculated light curve properties are also not sensitive to this choice (assuming a constant width for all shells, $\dR_0=\rm const$, gives a similar result).
Relative to \citet{Matsumoto&Metzger2022}, the updates of our current model include: (1) the diffusion time calculated as a full radial integral (Eq.~\ref{eq:t_diff}) instead of using a local estimate; (2) inclusion of the early-time suppression factor (Eq.~\ref{eq:suppression}).

For the assumed mass profile, the diffusion time is 
\begin{align}
t_{\rm d}&\simeq\frac{\beta}{\beta+2}\frac{\kappa M(>v)}{4\pi cR} \nonumber \\
&\underset{\beta = 7.6}\simeq{240\, \rm d\,}\left(\frac{M(>v)}{\Msun}\right)\left(\frac{R}{10^{3}R_{\odot}}\right)^{-1}\ ,
    \label{eq:t_diff_erpt}
\end{align}
where the second equality assumes a constant electron scattering opacity $\kappa=\kappa_{\rm es}\simeq0.32\,\rm cm^2\,g^{-1}$ (hereafter adopted to other analytic estimates unless otherwise specified).  The total ejecta kinetic energy is given by
\begin{align}
E_{\rm kin}&=\int_{v_{\rm ej}}\frac{dM}{dv}\frac{v^2}{2}dv\simeq\frac{\beta}{2(\beta-2)}M_{\rm ej}v_{\rm ej}^2
	\nonumber\\
\underset{\beta = 7.6}\simeq&5.2\times10^{47}{\,\rm erg\,}\biggl(\frac{M_{\rm ej}}{\Msun}\biggl)\biggl(\frac{v_{\rm ej}}{v_{\rm esc}}\biggl)^2\biggl(\frac{M_\star}{10\,\Msun}\biggl)\biggl(\frac{R_0}{10^2\,\Rsun}\biggl)^{-1}\ .
\end{align}
Based on the physical requirement that the total energy radiated by SN precursors $E_{\rm pre} \sim 10^{47}-10^{48}$ erg (Table \ref{table:obs}) not exceed $E_{\rm kin}$, we note that the minimum velocity must obey $v_{\rm ej}\gtrsim 340\,\kms\,(E_{\rm pre}/10^{48}{\,\rm erg\,})^{1/2}(M_{\rm ej}/\Msun)^{-1/2}$.

\subsection{Wind Scenario}
\label{sec:wind_setup}

In the steady wind scenario, the precursor emission is generated by a mass outflow from the star with an assumed constant mass-loss rate $\dot{M}$ and velocity $v_{\rm w}$.  We begin our calculation at the sonic radius $R_{\rm s}$, where the thermal energy flux of the wind is comparable to its kinetic power,
\begin{align}
\dot{E}_{\rm w} &=\frac{1}{2}\dot{M}v_{\rm w}^2\simeq1.2\times10^{40}\,\lumi
    \nonumber\\
&\times\biggl(\frac{\dot{M}}{\smyr}\biggl)\biggl(\frac{v_{\rm w}}{v_{\rm esc}}\biggl)^{2}\biggl(\frac{M_\star}{10\,\Msun}\biggl)\biggl(\frac{R_{\rm s}}{10^{2}\,\Rsun}\biggl)^{-1}\ ,
    \label{eq:wind_luminosity}
\end{align}
where we have scaled the wind velocity to the escape speed $v_{\rm esc}$ at the sonic radius (Eq.~\ref{eq:v_esc}, but replacing $R_0$ with $R_{\rm s}$).

To treat the problem within the same framework as the eruption case, the wind is divided into shells of equal mass $M = \dot{M}\dt$, radius $R=v_{\rm w}(t-t_0)$, width $\dR = v_{\rm w} \dt$, and volume $V=4\pi R^2\dR$, where $t_0$ is the time a given shell is released relative to the start of the wind ($t = 0$) and $\dt$ is the fixed time interval separating the ejection of successive shells.  The procedure to calculate the light curve from each shell follows that of the eruption scenario, except that the suppression factor (Eq.~\ref{eq:suppression}) is now given by
\begin{align}
S(t)=\frac{\left[e^{t_{\rm d}/(t-t_0)}-1\right]}{t_{\rm d}/(t-t_0)} \ .
\end{align}
While the diffusion timescale is calculated by Eq.~\eqref{eq:t_diff}, its estimate is given by
\begin{align}
t_{\rm d}&\simeq\frac{\kappa \dot{M}}{4\pi cv_{\rm w}} \simeq32{\,\rm d\,}\biggl(\frac{\dot{M}}{\Msun\,\rm yr^{-1}}\biggl)\times \nonumber \\
&\biggl(\frac{v_{\rm w}}{v_{\rm esc}}\biggl)^{-1}\biggl(\frac{M_\star}{10\,\Msun}\biggl)^{-1/2}\biggl(\frac{R_{\rm s}}{10^2\,\Rsun}\biggl)^{1/2}\ ,
\label{eq:tdwind}
\end{align}
which follows by taking the density as $\rho=\dot{M}/4\pi r^2v_{\rm w}$ (for a wind density profile $\rho \propto r^{-2}$, the diffusion time receives roughly equal contributions from all decades in radius; e.g., \citealt{Chevalier&Irwin11}).

In summary, the wind scenario is described by three parameters: wind mass-loss rate $\dot{M}$, wind velocity $v_{\rm w}$ (equivalently, sonic point $R_{\rm s}$ for a given progenitor mass $M_{\star}$), and the duration of the wind prior to the SN explosion $t_{\rm w}$.  Rather than define the latter as the entire duration of mass-loss prior to the explosion, $t_{\rm active}$, we define $t_{\rm w}$ as the more limited duration over which the luminosity has reached its roughly constant value, i.e. $t_{\rm w} = t_{\rm active} - t_{\rm rise},$ where $t_{\rm rise}$ is the light curve rise-time (see below, Sec.~\ref{sec:wind_results}).
  
What range of wind properties are expected physically?  \citet{Quataert+2016} show that energy injection at some radius $R_{\rm in}$ below the stellar surface at a super-Eddington rate $\dot{E} \gg L_{\rm Edd}$ drives a continuum radiation pressure-driven outflow.  For sufficiently high values of $\dot{E} \gg \dot{E}_{\rm Q}$, most of the injected power is eventually converted into the kinetic energy of the wind, i.e. $\dot{E}_{\rm w} \simeq \dot{E}$, where the threshold power is given by
\begin{align}
\dot{E}_{\rm Q} &\equiv f^{5/2}\frac{M_{\rm env}}{M_{\star}}\frac{v_{\rm esc}^{5}}{G}\simeq 2\times 10^{40}\,\lumi
    \nonumber \\
&\times\left(\frac{f}{0.3}\right)^{5/2}\left(\frac{M_{\rm env}}{10^{-2}M_{\star}}\right)\left(\frac{M_{\star}}{10\,\Msun}\right)^{5/2}\left(\frac{R_{\rm in}}{10^{2}\,\Rsun}\right)^{-5/2}.
\label{eq:EdotQ}
\end{align}
Here, $M_{\rm env} \lesssim 10^{-3}-10^{-2}\,M_{\star}$ is the mass of the stellar envelope above the energy injection point up to the sonic radius and $f \sim 0.1-1$ is a dimensionless parameter that depends on the envelope structure (\citealt{Quataert+2016}; their Fig. 11).  
Thus, we typically have $\dot{E}_{\rm Q} \sim 10^{39}-10^{41}{\,\rm erg\,s^{-1}}\sim 1-100\,L_{\rm Edd},$ where the Eddington luminosity
\begin{align}
L_{\rm Edd}&=\frac{4\pi G M_\star c}{\kappa_{\rm es}}\simeq1.6\times10^{39}{\,\lumi\,}\left(\frac{M_{\star}}{10\,\Msun}\right).\ 
\end{align}
In the limit $\dot{E} \gg \dot{E}_{\rm Q}$, \citet{Quataert+2016} show that the terminal velocity and mass-loss rate of the wind are related according to (their Eqs. 20, 26):
\begin{align}
v_{\rm w}&\simeq 1.7\left(\frac{M_{\star}}{M_{\rm env}}G\dot{E}\right)^{1/5} \simeq 430\,{\rm km\,s^{-1}}
\nonumber\\
&\times\left(\frac{M_{\rm env}}{10^{-3}\,M_{\star}}\right)^{-1/5}\left(\frac{M_{\star}}{10\,\Msun}\right)^{1/5}\left(\frac{\dot{E}}{100L_{\rm Edd}}\right)^{1/5}\ ,
    \label{eq:vQ}\\
\dot{M} &\simeq  0.15\,\frac{v_{\rm w}^{3}}{G}\frac{M_{\rm env}}{M_{\star}} \nonumber \\
&\simeq 0.96\,M_{\odot}{\rm \, yr^{-1}}\,
\left(\frac{M_{\rm env}}{10^{-3}\,M_{\star}}\right)\left(\frac{v_{\rm w}}{300\,{\rm km\,s^{-1}}}\right)^{3}.
\label{eq:MdotQ}
\end{align}
The lower range of ejecta velocities inferred from SN precursor observations, $v_{\rm w} \sim 100-500$ km s$^{-1}$ (Table \ref{table:obs}), can be obtained for $\dot{E} \sim 1-100\,L_{\rm Edd} \gtrsim \dot{E}_{\rm Q}$ and $M_{\rm env} \sim 10^{-4}-10^{-2}\,M_{\star}$, corresponding to range of $\dot{M} \sim 1-100\,M_{\odot}$ yr$^{-1}$.  On the other hand, the higher end of the observed precursor velocity $\gtrsim 10^3\,\kms$ would require both a small envelope mass $M_{\rm env}\lesssim 10^{-5}\,M_\star$ and large injection luminosity $\dot{E}\gtrsim 10^3\,L_{\rm Edd}$.

Although $v_{\rm w}$ represents the final velocity of the wind, \citet{Quataert+2016} found that the initial expansion rate of the overlying envelope material, after the onset of heating, is significantly slower than $v_{\rm w}$.  This rising velocity inevitably leads to shock interaction between the wind and the overlying envelope material.  Additional heating from such shocks may boost the transient luminosity compared to the constant-velocity wind modeled here (see Sec.~\ref{sec:shocks}).

\section{Results}\label{sec:result}

We now summarize the light curve results, separately in the eruption and wind scenarios.  

\subsection{Eruption Scenario}
\label{sec:eruption_results}

\begin{figure}
\begin{center}
\includegraphics[width=85mm, angle=0,bb=0 0 278 219]{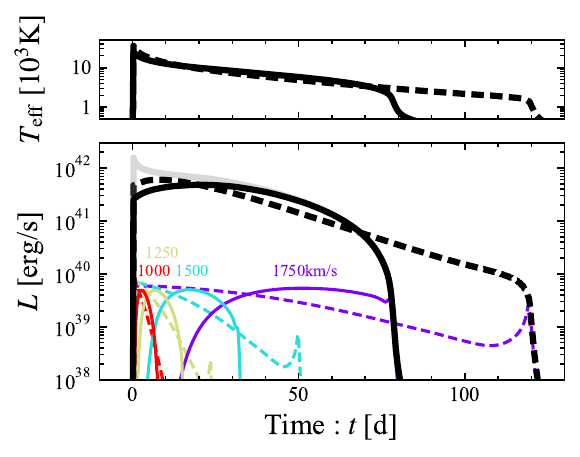}
\caption{Example SN precursor light curve calculation in the eruptive mass-loss scenario.  A solid black curve shows the optical light curve (bottom) and evolution of effective temperature (top) for a fiducial case with ejecta mass $M_{\rm ej}=\Msun$, initial radius $R_0 = R_\star=10^3\,\Rsun$, and minimum ejecta velocity $v_{\rm ej}=1000\,\kms$ for an assumed stellar mass $M_\star = 10\, \Msun$.  The optical luminosity is calculated by excluding radiation of frequency $>10^{15}$ Hz from the bolometric light curve (gray curve).  Colored solid lines show the luminosity contributed by single shells with velocity $v=1000\,\kms$ to $1750\,\kms$ (separated by $250\,\kms$) and shell width of $dv=2\,\kms$.  Dashed curves show an otherwise identical calculation calculated following the original model of \citet{Matsumoto&Metzger2022}, which neglects the early-time suppression factor (Eq.~\ref{eq:suppression}) and uses a local estimate for the radial optical depth.}
\label{fig:lc_eruption_example}
\end{center}
\end{figure}

\begin{figure}
\begin{center}
\includegraphics[width=85mm, angle=0,bb=0 0 286 257]{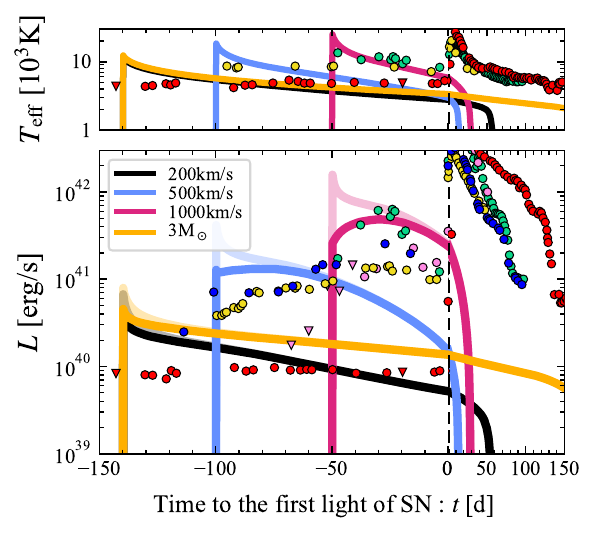}
\caption{Same as in Fig.~\ref{fig:lc_eruption_example} but showing light curves in the eruption scenario for different ejecta parameters, and compared with observed SN precursors (Fig.~\ref{fig:obs}).  Blue, black, and red curves show models with a common ejecta mass $M_{\rm ej}=\Msun$ but different ejecta velocities $v_{\rm ej}=200$, 500, and $1000\,\kms$, respectively. A yellow curve shows a model with $v_{\rm ej}=200\,\kms$ but larger $M_{\rm ej}=3\,\Msun$. Lighter shaded curves denote the bolometric luminosity. The stellar mass $M_\star=10\,\Msun$ and radius $R_0=10^3\,\Rsun$ are fixed for all models.  The time of the eruption prior to the SN is set to $t_{\rm erupt}$ = 140 days (except for high-velocity models shown with blue and red curve, for which $t_{\rm erupt} = 100$ and $50$ d) to roughly match the onset of the observed precursor emission.  The dependence of the predicted plateau luminosity and duration on the ejecta properties broadly agree with the Popov analytic scalings (Eqs.~\ref{eq:L_popov}, \ref{eq:t_popov}).}
\label{fig:lc_eruption}
\end{center}
\end{figure}

Figure~\ref{fig:lc_eruption_example} shows example light curves in the eruption scenario, calculated for: $M_{\rm ej}=1\,\Msun$, $M_\star=10\,\Msun$, $R_0=R_\star=10^3\,\Rsun$, and $v_{\rm ej} = 1000\,\kms$.  The chosen values of stellar mass and radius are representative of RSG progenitors, while the velocity is motivated by those observed spectroscopically. The ejecta mass is chosen as needed to explain observed precursor luminosities (see Fig.~\ref{fig:lc_eruption}).
For these parameters, the ejecta is radiation-pressure-dominated and the light curve is powered by the escape of the initial internal energy.  Hydrogen begins to recombine once  the temperature decreases to a value $T_{\rm ion}\sim10^{4}\,\rm K$, and the resulting sharp drop in opacity creates a well-defined recombination front.  As the front recedes through the ejecta shell, this shapes a characteristic plateau in the light curve in the same way as in Type IIP SNe \citep{Arnett1980,Popov1993,Kasen&Woosley2009}.  The top panel shows the effective temperature, $T_{\rm eff}=\left[L/(4\pi \sigma_{\rm SB}R_{\rm ph}^2)\right]^{1/4}$, where $\sigma_{\rm SB}$ is the Stefan-Boltzmann constant and the photosphere radius $R_{\rm ph}$ is defined by $\tau=\int_{R_{\rm ph}}\kappa\rho dr=2/3$.

Just after the eruption, the effective temperature is high and the peak of the assumed blackbody spectrum is in the UV range.  We approximate the optical-wavelength light curve (black curve) by subtracting the UV luminosity (defined as radiation with frequency $\nu>10^{15}\,\rm Hz$) from the bolometric luminosity (gray curve).  Colored solid lines denote the single-shell light curves for different velocities ($v=1000\,\kms$ to $1750\,\kms$, corresponding to colors from purple to red).  As a result of the suppression factor (Eq.~\ref{eq:suppression}), low velocity shells do not contribute significantly to the total luminosity at early times, until their diffusion times become shorter than the expansion times.

In order to quantify the impact of new features of the light curve model introduced in this paper, a dashed line show the calculation performed with the original model of \citet{Matsumoto&Metzger2022}, which does not include the early-time suppression effect (Eq.~\ref{eq:suppression}) and used a local estimate for the radial optical depth rather than a full integral taken over the external shells.  The new calculation light curve exhibits a shallower decay, but its overall luminosity and duration, do not change appreciably from the \citet{Matsumoto&Metzger2022} model.

Due to the close similarity of the physical conditions in the precursor ejecta to that of Type IIP SNe, the characteristic luminosity and duration of the light curve can be estimated using the analytic expressions from \citet{Popov1993}:
\begin{align}
L_{\rm Popov}&=2.0\times10^{41}{\,\rm erg\,s^{-1}\,}\left(\frac{M_{\rm ej}}{\Msun}\right)^{1/3}
    \nonumber\\
&\times\left(\frac{R_0}{10^3\,\Rsun}\right)^{2/3}\left(\frac{v_{\rm ej}}{10^3\,\rm km\,s^{-1}}\right)^{5/3}\ ,
	\label{eq:L_popov}\\
t_{\rm Popov}&=74{\,\rm d\,}\left(\frac{M_{\rm ej}}{\Msun}\right)^{1/3}\left(\frac{R_0}{10^3\,\Rsun}\right)^{1/6}\left(\frac{v_{\rm ej}}{10^3\,\rm km\,s^{-1}}\right)^{-1/3}\ ,
	\label{eq:t_popov}
\end{align}
where we have approximated the mean ejecta velocity as $v_{\rm ej}$ and have adopted a normalization calibrated from the radiative transfer simulations of \cite{Sukhbold+2016} (see also \citealt{Blagorodnova+2021}).

Figure~\ref{fig:lc_eruption} shows light curves in the eruption scenario, calculated for different variations of the ejecta parameters relative to the fiducial case in Fig.~\ref{fig:lc_eruption_example}.  The resulting changes in the plateau emission properties roughly agree with the analytic scaling relations in Eqs.~\eqref{eq:L_popov}, \eqref{eq:t_popov}. For example, increasing the ejecta mass increases the duration and luminosity of the plateau, while increasing the minimum velocity increases the plateau luminosity but shortens its duration.  

For comparison in Fig.~\ref{fig:lc_eruption} we show the light curves of observed SN precursors (Fig.~\ref{fig:obs}).  In order to roughly match the observations, we set the time of the eruption prior to core collapse to $t_{\rm erupt} = 140$, $100$, and $50$ days for $v_{\rm ej} = 200\,\kms$ (black and yellow), $500\,\kms$ (blue), and $1000\,\kms$ (red), respectively. Before exploring the detailed model parameters required to fit the observations, we first note that$-$broadly speaking$-$the luminosities predicted in the outburst model, $L_{\rm pre}\sim10^{41}\,\rm erg\,s^{-1}$, match those of observed precursors.  In detail, our outburst model predicts light curves which decay monotonically following the initial peak, in conflict with observed precursor emission (in particular, SN 2015bh and 2016bdu), which instead rises for a few months leading up to the SN.  However, this discrepancy may at least in part arise from some of the simplifying assumptions of our model, such as the velocity profile and initial shell radii; indeed, the radiation hydrodynamic simulations of precursor outbursts by \cite{Dessart+2010} predict rising light curves.\footnote{Adopting the same ejecta parameters as \cite{Dessart+2010}, our calculation predicts a plateau luminosity and duration that agree with their numerical results to within a factor of a few.}

\begin{figure*}
\begin{center}
\includegraphics[width=135mm, angle=0,bb=0 0 499 642]{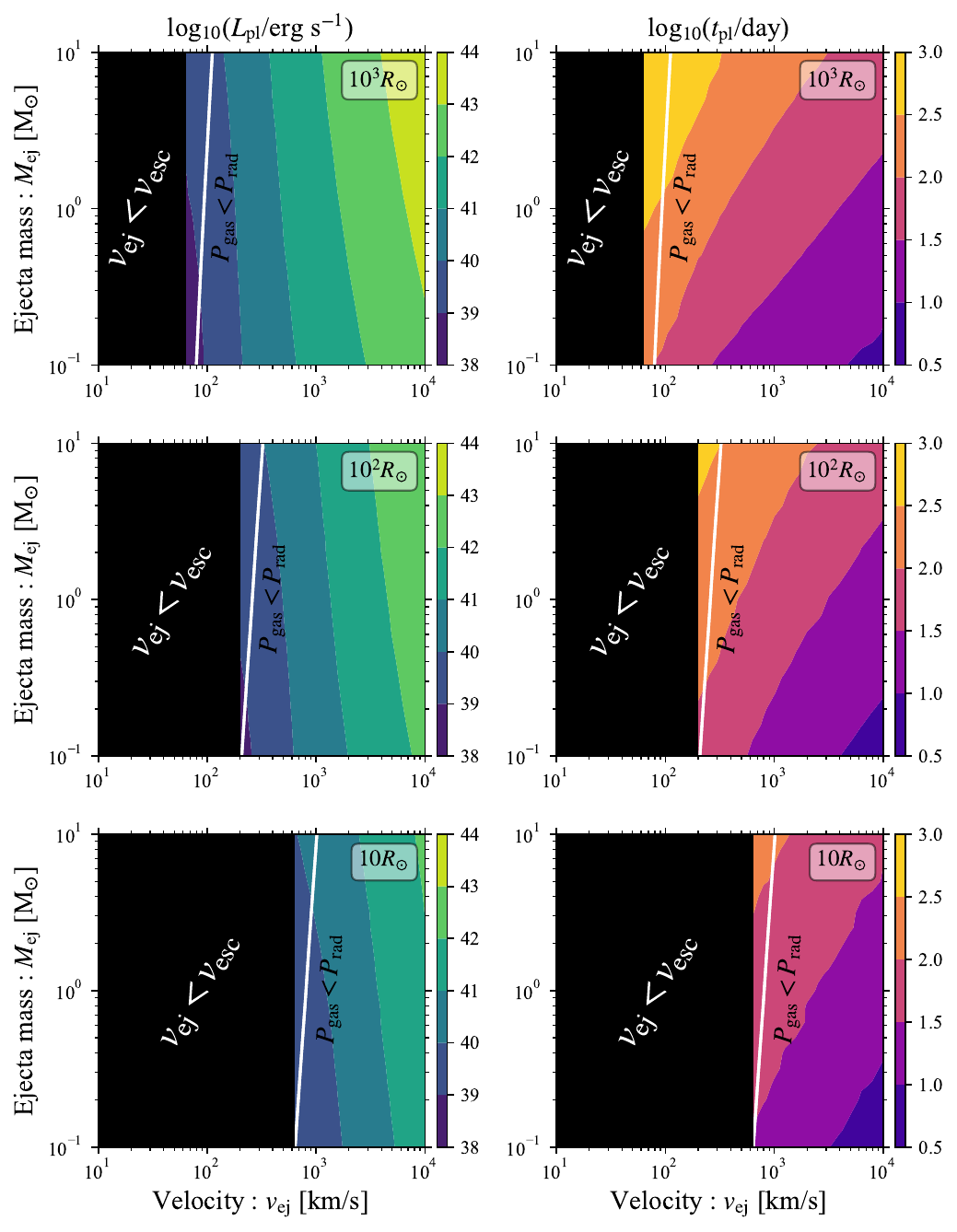}
\caption{Contours of the characteristic optical precursor luminosity $L_{\rm pl}$ (after subtracting the UV luminosity; left column) and duration $t_{\rm pl}$ (right column) associated with eruptive mass-loss events in the space of ejecta velocity and mass, for different initial radii $R_0=10^3$, $10^2$, and $10\,\Rsun$ (from top to bottom). Black shaded regions are excluded because the ejecta velocity is smaller than the surface escape speed for a star of mass $M_{\star} = 10\,\Msun$ and radius $R_{\star} \le R_0$.  To the right of the white contours radiation-pressure exceeds gas-pressure at the initial radius.}
\label{fig:dist_erpt}
\end{center}
\end{figure*}

Figure~\ref{fig:dist_erpt} shows an estimate of the plateau luminosity $L_{\rm pl}$ and duration $t_{\rm pl}$, calculated from a large grid of models in the space of ejecta mass $M_{\rm ej}$ and ejecta velocity $v_{\rm ej}$ for different assumptions about initial (and progenitor) radii $R_0=R_\star = 10$, $10^{2}$, and $10^{3}\,\Rsun$. Here, we define the duration $t_{\rm pl}$ as that over which 90\% of the total energy is radiated, $0.9E_{\rm pl}=\int^{t_{\rm pl}}Ldt$, where $E_{\rm pl}=\int Ldt$.  The average luminosity is then defined as $L_{\rm pl} = E_{\rm pl}/t_{\rm pl}$.  Note that the model-predicted plateau duration $t_{\rm pl}$ only represents an upper limit on the observed precursor duration if the SN explosion occurs before the end of the eruption emission phase.  However, due to the relatively flat shape of the predicted light curve, the model-predicted value of $L_{\rm pl}$ will still be comparable to the time-averaged observed luminosity $L_{\rm pl}$, even when the eruption emission is prematurely terminated by the SN.  

The distributions of plateau luminosity and duration roughly follow those predicted by Eqs.~\eqref{eq:L_popov} and \eqref{eq:t_popov}, namely $M_{\rm ej}\propto v_{\rm ej}^{-5}$ and $M_{\rm ej}\propto v_{\rm ej}$, for fixed values of $L_{\rm Popov}$ and $t_{\rm Popov}$, respectively.  For lower ejecta velocities, comparable to the RSG surface escape speed $\sim 10-100$ km s$^{-1}$, gas pressure instead of radiation pressure dominates in the ejecta and other assumptions of the Popov analytic estimates (e.g., the neglect of hydrogen recombination energy) are violated \citep{Matsumoto&Metzger2022}.  On the other hand, for ejecta masses $\sim\Msun$, velocities $\sim 10^3\,\rm km\,s^{-1}$, and radii close to the surface of the RSG progenitors ($R_0 \sim R_\star \sim 10^{3}\,\Rsun$), the predicted transient properties are consistent with the observed precursors.  The high required ejecta velocity exceeding the surface escape speed by factors $\gtrsim 10$ may point to energy deposition deep within the star $\sim10\,\Rsun$.  To reproduce the observed precursors with $\sim10^{41}\,\rm erg\,s^{-1}$ and $\sim10^3\,\rm km\,s^{-1}$ from more compact progenitor stars such as blue supergiants ($R_\star\sim10\,\Rsun$, see the bottom panel in Fig.~\ref{fig:dist_erpt}), would require more massive ejecta $\gtrsim 3\,\Msun$.

\subsection{Wind Scenario}
\label{sec:wind_results}

\begin{figure}
\begin{center}
\includegraphics[width=85mm, angle=0,bb=0 0 281 256]{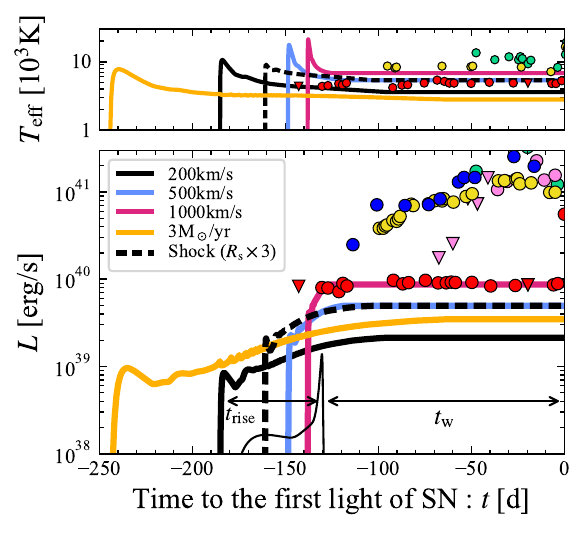}
\caption{Optical light curves of SN precursor emission in the wind mass-loss scenario for progenitor mass $M_\star=10\,\Msun$. A thick black curve shows the fiducial case with wind mass-loss rate $\dot{M}= 1 \Msun\,\rm yr^{-1}$ and wind velocity $v_{\rm w}=200\,\kms$ ($R_{\rm s}=10^2\,\Rsun$).  The wind duration is set to $t_{\rm w}=130$ d to match the SN2020tlf precursor duration (shown for comparison red points, along with the other precursors labeled as in Fig.~\ref{fig:obs}). A thin black curve shows the light curve from a single shell of thickness $\dt=4\,\rm d$ for the fiducial model, whose duration$-$being similar for all shells$-$controls the rise-time of the total light curve. Colored lines show the light curve that results by increasing the wind mass-loss rate or speed relative to the fiducial model (while fixing $t_{\rm w}$). A black dashed curve shows the effect of tripling the ``sonic'' radius at which the wind thermal energy is initialized, to crudely mimic the effects of re-heating of the ejecta by internal shocks (Sec.~\ref{sec:shocks}).}
\label{fig:lc_wind}
\end{center}
\end{figure}

Figure \ref{fig:lc_wind} shows examples of the light curve and effective temperature evolution in the wind mass-loss scenario, for a fixed progenitor mass $M_\star=10\,\Msun$.
A thick black curve depicts the chosen fiducial model, with wind mass-loss rate $\dot{M}=\smyr$ and speed $v_{\rm w}=200\,\kms$ (corresponding to $R_{\rm s}=100\,\Rsun$, Eq.~\ref{eq:v_esc}). This mass loss rate is broadly motivated by observed Type IIn SNe, though is somewhat higher than the values typically inferred (albeit usually on larger radial scales; e.g., \citealt{Smith2014}).
As described in Sec.~\ref{sec:wind_setup}, the total light curve is obtained by adding those from successive single shells (an example of which is shown by a thin black curve in Fig.~\ref{fig:lc_wind}).  Since the total luminosity approaches a steady-state on the timescale roughly set by the characteristic duration of the single shell emission, we denote the latter $t_{\rm rise}$ and calculate its value in the same way as the eruption model at the stationary state (i.e., it is defined by the timescale over which 90\% of energy is radiated). The rise-time is roughly equal to that over which photons diffuse out from the shell, $t_{\rm d}$ (Eq.~\ref{eq:tdwind}).

The single-shell light curve exhibits a gradual rise followed by sharp peak, powered by the escape of energy that accompanies a rapid decrease in the photon diffusion timescale at hydrogen recombination. For the fiducial model, radiation pressure dominates (marginally) over gas pressure during the recombination phase, rendering the contribution of recombination energy to the radiated energy small.  The luminosity reached during the steady plateau phase is roughly given by the product of the single shell's characteristic luminosity calculated in the same way as in the eruption model (dividing the radiated energy by $t_{\rm rise}$), by the number of shells ejected during the rise phase $\simeq t_{\rm rise}/\dt$. We remark that the rise-time $t_{\rm rise}(\sim t_{\rm d})$ is independent of the choice of shell thickness $v\dt$ because it does not enter in Eq.~\eqref{eq:tdwind}.

Colored curves in Fig.~\ref{fig:lc_wind} show additional light curve calculations varying the wind mass-loss rates and velocities from the fiducial model.  Because the rise-time is roughly set by the diffusion timescale $t_{\rm d} \propto \dot{M}/v_{\rm w}$ (Eq.~\ref{eq:tdwind}), the steady-state luminosity is achieved faster for higher wind velocities.  Similarly, winds with higher mass-loss rates take longer to reach steady-state.

As an aside, we remark that for winds with low velocities and/or high-$\dot{M}$, our calculation overestimates the photosphere radius and hence underestimates (by less than a factor of $\lesssim$ 2) the effective temperature of the emission.  As a result of the shallow wind density profile $\rho \propto r^{-2}$, material ahead of the recombination front (which quickly cools to $\lesssim10^3\,\rm K$) can contribute significantly to the optical depth through molecular absorption.  However, at such low temperatures, our analytic opacity law (which adopts a constant value $\kappa_{\rm mol} = 0.1Z \sim 10^{-3}$ cm$^{2}$ g$^{-1}$) overestimates the true molecular opacity.  Unless significant dust formation occurs to boost the opacity further \citep{Freedman+2008,Pejcha+2016}, the true photosphere location should therefore coincide with the recombination front, at somewhat smaller radii than our model predicts.

\begin{figure*}
\begin{center}
\includegraphics[width=165mm, angle=0,bb=0 0 505 534]{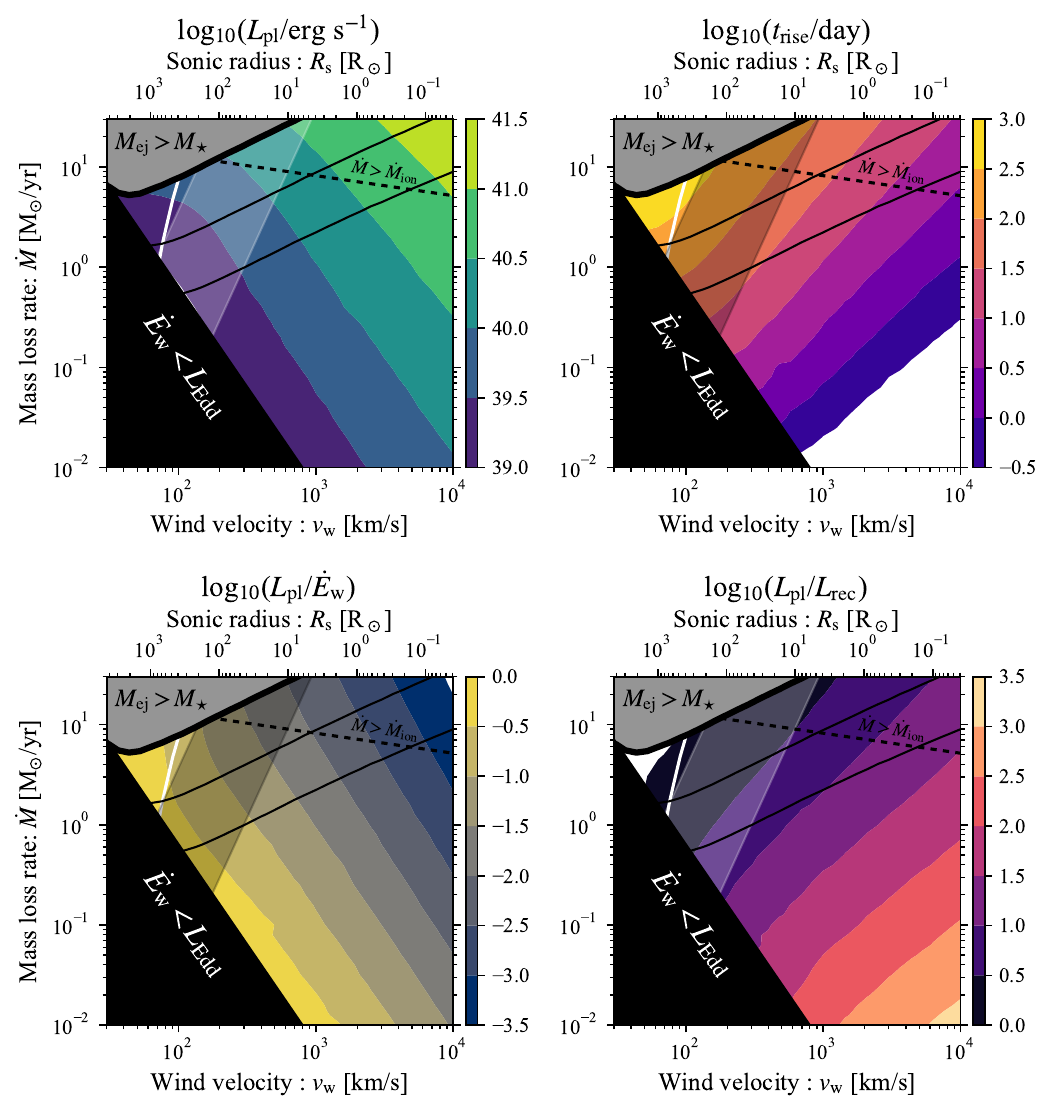}
\caption{{\it Top:} Steady-state $L_{\rm pl}$ luminosity (left panel) and rise-time $t_{\rm rise}$ (right panel) of SN precursor emission from wind-like mass-ejection as a function of the wind mass-loss rate $\dot{M}$ and velocity $v_{\rm w}$ (equivalently, sonic radius for an assumed progenitor star mass $M_{\star} = 10\,\Msun$; top horizontal axis).  Black and gray shaded regions are excluded based on the wind kinetic luminosity being sub-Eddington and the total ejecta mass during the rise-time exceeding the progenitor star mass (Eq.~\ref{eq:Mdot_max1}).  The other solid black lines correspond to $M_{\rm ej} = 0.1M_{\star}$ and $0.01M_{\star}$.  Thin shaded regions represent the mass-loss rate $\dot{M}(v_{\rm w})$ predicted by \citet{Quataert+2016} (Eq. \ref{eq:MdotQ}), for a range of envelope masses $M_{\rm env}/M_{\star}=0.001-0.1$.  A black dashed line shows the critical mass-loss rate $\dot{M}_{\rm ion}$ (Eq.~\ref{eq:Mdotcri} for $T_{\rm ion} = 10^{4}$ K) above which the drop in opacity from hydrogen recombination determines the transient rise-time.  Radiation pressure exceeds gas pressure at the sonic point to the right of the white line, i.e. across most of the parameter space. {\it Bottom:} Same as top panels, but showing the ratios of steady-state $L_{\rm pl}$ luminosity to the wind kinetic power, $\dot{E}_{\rm w}$ (Eq.~\ref{eq:wind luminosity}; left panel), and to the maximum recombination luminosity, $L_{\rm rec}$ (Eq.~\ref{eq:Lrec}; right panel).}
\label{fig:dist_wind}
\end{center}
\end{figure*}

\begin{figure}
\begin{center}
\includegraphics[width=85mm, angle=0,bb=0 0 277 216]{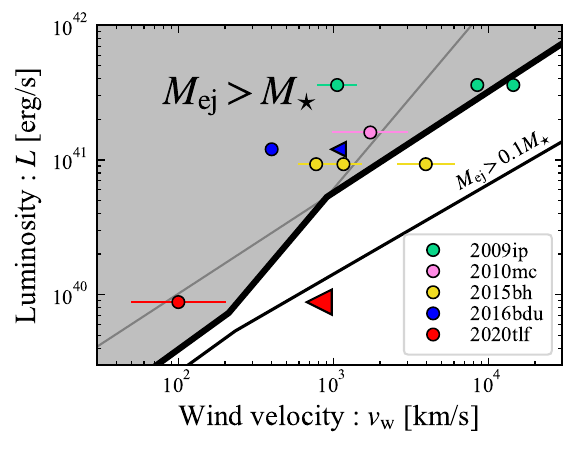}
\caption{The steady wind scenario is challenged to explain high luminosities of SN precursor emission for reasonable ejecta masses.  Black lines shows the maximum steady-state luminosity in the wind scenario as a function of wind velocity so as not to overproduce the total wind ejecta mass $M_{\rm ej}$ (Eqs.~\ref{eq:Lmax1}, \ref{eq:Lmax2}); specifically, the gray shaded region is excluded by the requirement $M_{\rm ej}>M_\star=10\,\Msun$ (see text for details).  Colored circles represents observed SN precursor luminosities and velocities (Table \ref{table:obs}).  Left-directed triangles show an estimate of the wind velocity ($v_{\rm engulf}$; Eq.~\ref{eq:vengulf}) invisible to flash spectroscopy observations (too slow wind material would be overtaken by the SN shock prior to the observations).}
\label{fig:Lmax_wind}
\end{center}
\end{figure}

The top panels of Fig.~\ref{fig:dist_wind} show the rise-time $t_{\rm rise}$ (upper right panel) and steady-state luminosity $L_{\rm pl}$ (upper left panel) as a function of the wind velocity $v_{\rm w}$ and mass-loss rate $\dot{M}$, for fixed progenitor mass $M_\star=10\,\Msun$.  Regions of the phase space for which the wind kinetic energy is below the Eddington luminosity of the star are blacked out, based on the expectation that the star would respond to sub-Eddington energy deposition by driving convection or expanding, rather than give rise to significant mass-loss (e.g., \citealt{Quataert+2016}).  Likewise, regions for which the total mass-loss in wind ejecta experienced over the light curve rise-time, exceeds the total stellar mass (i.e., $M_{\rm ej} \simeq \dot{M} t_{\rm rise} > M_{\star}$) are shaded gray. As shown in Fig.~\ref{fig:Lmax_wind}, this condition imposes severe limits on the wind scenario. White shaded regions denote the mass-loss rate $\dot{M}(v_{\rm w})$ predicted by \citet{Quataert+2016} (Eq.~\ref{eq:MdotQ}) for a range of heated envelope masses $M_{\rm env} \sim 10^{-3}-10^{-1}M_{\star}$.

In Appendix \ref{sec:appendix} we derive analytic estimates of the steady-state luminosity as a function of the wind properties.  There are two regimes to consider, depending on whether the ejecta is still ionized or has begun to recombine by the radius of emission (as dictated by whether $\dot{M}$ is below or above, respectively, a critical value $\dot{M}_{\rm ion}$; Eq.~\ref{eq:Mdotcri}).  For $\dot{M} < \dot{M}_{\rm ion}$ we find $L \propto \dot{M}^{1/3}v_{\rm w}^{2/3}$ (Eq.~\ref{eq:LlowMdot}; see also \citealt{Quataert+2016,Shen+2016}), while for $\dot{M} > \dot{M}_{\rm ion}$ we find $L \propto \dot{M}^{3/4}v_{\rm w}^{3/4}$ (Eq.~\ref{eq:LhighMdot}).  These dependencies roughly agree with our findings in Figs.~\ref{fig:lc_wind}, \ref{fig:dist_wind}.

The bottom left panel of Fig.~\ref{fig:dist_wind} shows the ratio of steady-state transient luminosity to the wind kinetic power $L_{\rm pl}/\dot{E}_{\rm w}$, i.e., a measure of the wind's radiative efficiency.  The efficiency decreases as the wind power $\dot{E}_{\rm w} \propto \dot{M}v_{\rm w}^{2}$ increases, roughly according to $L_{\rm pl}/\dot{E}_{\rm w} \propto (\dot{E}_{\rm w}/L_{\rm Edd})^{-2/3}$ (see Eq.~\ref{eq:LlowMdot}).  Larger values of $\dot{M}$ push the diffusion radius ($\sim v_{\rm w}t_{\rm d}$) outwards, while larger $v_{\rm w}$ shrink the sonic radius, both effects of which result in greater adiabatic losses prior to where radiation can escape from the wind.  

The bottom right panel of Fig.~\ref{fig:dist_wind} shows the ratio of $L_{\rm pl}$ to maximum recombination luminosity,
\begin{align}
L_{\rm rec} =\frac{\varepsilon_{\rm H}X\dot{M}}{\MP} \simeq6.1\times10^{38}{\,\lumi\,}\biggl(\frac{\dot{M}}{\smyr}\biggl)\ ,
    \label{eq:Lrec}
\end{align}
where $\varepsilon_{\rm H}=13.6\,\rm eV$ is the recombination energy per hydrogen atom of mass $\MP$ and $X=0.74$ is the hydrogen mass fraction of the ejecta at solar metallicity.  Unlike in the case of most stellar merger transients \citep{Metzger+2021,Matsumoto&Metzger2022}, the radiated luminosity generally exceeds the recombination luminosity insofar as radiation pressure dominates over gas pressure (as satisfied to the right of the solid white line in Fig.~\ref{fig:dist_wind}). 

For $\dot{M} < \dot{M}_{\rm ion},$ we invert Eqs.~\eqref{eq:tdwind}, \eqref{eq:LlowMdot} to obtain the wind mass-loss rate and velocity in terms of the observed precursor luminosity and rise-time\footnote{Constraining the rise-time is challenging in practice because current observations do not probe the earliest stages of the precursor emission (Fig.~\ref{fig:obs}); furthermore, the luminosity evolves slowly even before reaching steady-state, making it hard to cleanly distinguish the rising phase.}:
\begin{align}
\label{eq:invert1}
v_{\rm w}&\simeq940{\,\kms\,}\left(\frac{M_\star}{10\,\Msun}\right)^{-2/3}\left(\frac{t_{\rm rise}}{10{\,\rm d}}\right)^{-1/3}\left(\frac{L_{\rm pl}}{10^{40}{\,\rm erg\,s^{-1}}}\right)\ ,\\
\dot{M}&\simeq1.5{\,\smyr\,}\left(\frac{M_\star}{10\,\Msun}\right)^{-2/3}\left(\frac{t_{\rm rise}}{10{\,\rm d}}\right)^{2/3}\left(\frac{L_{\rm pl}}{10^{40}{\,\rm erg\,s^{-1}}}\right)\ .
\label{eq:invert2}
\end{align}
These expressions reveal that very large mass-loss rates $\dot{M} \gtrsim 1-10\,\smyr$ are required to explain the SN precursor luminosities, $L_{\rm pre} \sim 10^{40}-10^{41}$ erg s$^{-1}$.  In fact, a conservative upper limit can be placed on the steady-state plateau luminosity in the wind-scenario under the condition that total wind ejecta mass $M_{\rm ej}$ not exceed the progenitor star mass.  This translates into an upper limit on the wind mass-loss rate: $M_{\rm ej}=t_{\rm active}\dot{M}<M_\star$, where $t_{\rm active}$ is the duration of mass-loss episode (from the wind launch to the SN explosion). This timescale may be reasonably approximated by the longer of the (theoretical) light curve rise-time, $t_{\rm rise}$, or the observed precursor plateau duration, $t_{\rm pre}$.  For $t_{\rm rise}$, the maximal $\dot{M}$ and corresponding maximum steady-state luminosity $L_{\rm max}$ are derived in Appendix~\ref{sec:appendix}  (Eqs.~\ref{eq:Lmax1}, \ref{eq:Lmax2}). For $t_{\rm pre}$, the maximal luminosity follows by substituting $\dot{M} = \dot{M}_{\rm max} =M_\star/t_{\rm pre}$ into Eqs.~\eqref{eq:LlowMdot}, \eqref{eq:LhighMdot}.
We set the maximal luminosity by taking the smaller one calculated for $t_{\rm rise}$ and $t_{\rm pre}$.

This theoretical maximum luminosity is shown with a solid black line in Fig.~\ref{fig:Lmax_wind} as a function of $v_{\rm w}$ for $M_{\star}=10\,\Msun$ and a plateau duration $t_{\rm pre}=100\,\rm d$ close to those observed (Fig.~\ref{fig:obs}).  For lower wind speeds $v_{\rm w}\lesssim1000\,\kms$, the light curve rise-time exceeds the observed plateau duration $t_{\rm rise}>t_{\rm pre}$, placing the tighter limit on the maximal luminosity.  For comparison, colored symbols show the time-averaged luminosities and spectroscopically-measured outflow velocities for individual SN precursors (Table \ref{table:obs}). The precursor luminosities typically exceed the theoretical limit by at least a factor of a few, thus challenging the wind scenario.  

In precursor events for which flash-spectroscopy observations were conducted soon after the SN explosion, the measured CSM velocities from these observations ($v_{\rm obs}$ in Table \ref{table:obs}) are in two cases (SN 2016bdu and SN 2020tlf) so low the SN ejecta would have swept up any precursor-ejected material before the observations were taken (hence, the CSM velocities measured must arise from an earlier mass-loss episode prior to the precursor rise).  This hypothetical requirement on the measurable CSM velocity can be written: 
\begin{align}
&v_{\rm obs} \gtrsim v_{\rm engulf} \equiv v_{\rm SN}\frac{t_{\rm flash}}{t_{\rm pre}} \nonumber \\
&\simeq 1000\,\kms\,\left(\frac{v_{\rm SN}}{10^4{\,\kms}}\right)\left(\frac{t_{\rm flash}}{10{\,\rm d}}\right)\left(\frac{t_{\rm pre}}{100{\,\rm d}}\right)^{-1},
\label{eq:vengulf}
\end{align}
where $v_{\rm SN}$ and $t_{\rm flash}$ are the SN ejecta velocity and time of the flash-spectroscopy observation measured since the SN explosion, respectively.\footnote{Note that here we define the critical velocity not by the entire wind active period $t_{\rm active}$ but rather the observed duration $t_{\rm pre}$; this is justified because emission from the maximal CSM velocity material typically rises quickly $t_{\rm rise}\lesssim t_{\rm pre}$.} As a corollary, CSM with velocities $v_{\rm obs} > v_{\rm engulf}$ could in principle (if present) be observed in systems with flash spectroscopy observations, thus enabling an upper limit on $v_{\rm w} < v_{\rm engulf}$ based on the lack of detection.  This maximal observable velocity $v_{\rm engulf}$ are shown as left-facing arrows in Fig.~\ref{fig:Lmax_wind} for $v_{\rm SN}=10^4\,\kms$.  

Taken together, we conclude that only SN 2020tlf can be explained by the wind scenario for a reasonable ejecta mass $\lesssim0.1\,M_{\star}\sim\Msun$, and in this case only if the wind velocity (inaccessible to flash spectroscopy) were to obey $600\,\kms \lesssim v_{\rm w} \lesssim 1000\,\kms$. These values could be consistent with a RSG progenitor whose envelope mass outside of the implied sonic radius $R_{\rm s}\sim10\,\Rsun$ is on the order of $\sim\Msun$. Interestingly, these parameters are also consistent with the wind solution of \cite{Quataert+2016} for the envelope mass of $M_{\rm env}\sim10^{-4}-10^{-5}\,M_\star$. Given the implied small sonic radius of $R_{\rm s}\sim10\,\Rsun$, we can estimate the mass included between the energy injection and sonic radii by $M_{\rm env}\sim\dot{M}R_{\rm s}/v_{\rm w}\sim10^{-5}\,\Msun\sim10^{-4}\,M_\star$, consistent with the required value of $M_{\rm env}.$

\subsubsection{Shock Interaction}
\label{sec:shocks}

A potential mechanism to boost the luminosity of the wind is through shock interaction.  By reheating the wind ejecta well above the sonic surface, shocks reduce the adiabatic losses experienced before the flow reaches the diffusion radius where the remaining thermal energy is radiated (Eq.~\ref{eq:r_diff}).  One source of shock interaction are internal collisions within the wind itself, due to time variability in the wind speed.

Consider two shells with the same mass $M$ but different velocities.
The slower shell with velocity $v$ is ejected $\delta t$ before the faster one with velocity $(1+\delta_v)v$.
The latter one catches up with former one in a time $\delta t/\delta_v$ and they merger to form a single shell.  The collision radius is estimated by $R_{\rm col}\sim (2+1/\delta_v)R_{\rm s}$, where we have assumed both shells originate from the same radius $\sim R_{\rm s}$ separated by a time $\delta t\sim R_{\rm s}/v$ roughly equal to the dynamical time at the launching surface.  From momentum and energy conservation, the final velocity of the merged-shell is $(1+\delta_v/2)v$ and the energy dissipated by the collision is $M(v\delta_v)^2/2$.

The velocity difference $\delta_v$ has two competing effects on the radiated luminosity of the shocked wind.  While smaller values of $\delta_v$ increase the collision radius and thus reduce the adiabatic losses, such collisions also dissipate less energy than for larger $\delta_v$.  Adiabatic losses exterior to $R_{\rm col}$ reduce the observed luminosity by an amount $L \propto R_{\rm col}^{-2/3} \propto (2+1/\delta_v)^{-2/3}$ (Eq.~\ref{eq:wind luminosity}), while the boost in thermal energy $\propto (\delta_v)^2$; cases with the largest, order-unity fluctuations $\delta_v \sim 1$ therefore give the biggest luminosity boost.  

To demonstrate the potential effects of shock heating, we recalculate the fiducial wind scenario light curve for the (most luminous) case $\delta_v=1$ by artificially increasing the ``sonic'' radius $R_{\rm s}$, at which the wind thermal energy is initialized in our calculation, by a factor of three compared to the fiducial case.  As shown in Fig.~\ref{fig:lc_wind}, the radiated luminosity roughly doubles compared to the case without shock interaction.  Our simplified treatment of the effects of shock heating results in a smooth plateau-shaped light curve similar to the case without shocks.  In reality, however, the light curve should exhibit variability due to the stochastic nature of the heating, though this could be smoothed out by photon diffusion effects.  Most SN precursor light curves are not perfectly smooth (Fig.~\ref{fig:obs}), with the exception of SN 2020tlf (for which shock heating is anyways not strictly required in the wind scenario; Fig.~\ref{fig:Lmax_wind}). 

Another source of shock interaction is between the wind and pre-existing CSM ejected earlier from the star,\footnote{Hypothetical dense clumps bound to and accumulating around the progenitor may also play a role of the dense CSM \citep{Soker2021}.} or, equivalently, if the wind velocity rises in time.  In fact, \cite{Quataert+2016} found that shock interaction inevitably happens before the wind settles into a steady state.  After the onset of energy injection below the stellar surface, the stellar envelope is initially only weakly accelerated to form a slowly-expanding atmosphere.  Continuous heating decreases the density at the heating site, accelerating the material there to higher velocity (Eq.~\ref{eq:vQ}).  The high-speed wind then collides with the slowly-expanding atmosphere, generating shocks which can dominate the radiated luminosity, at least initially.  Although we defer a detailed study to future work, shock interaction due to an accelerating wind provide another way to increase the wind's radiative efficiency, particularly at early times.

\section{Precursor-Generated CSM}\label{sec:discussion}

The mass ejected from the progenitor star during the precursor phase becomes a source of CSM for shock interaction with the SN ejecta following the explosion.  The application of our precursor light curve models to individual events enables us to constrain their ejecta masses and velocities, which we can then check for consistency with the CSM properties inferred by modeling interaction signatures during the SN phase.

\subsection{Eruption scenario}
We have found (Sec.~\ref{sec:eruption_results}) that the eruption scenario can account for the luminosities and timescales of SN precursor emission for physically reasonable ejecta properties. The ejecta mass and initial radius in this scenario are obtained by inverting Eqs.~\eqref{eq:L_popov}, \eqref{eq:t_popov} in terms of the precursor luminosity $L_{\rm pl}$ and duration $t_{\rm pl}$:
\begin{align} 
M_{\rm ej}&\simeq6.7\,\Msun\left(\frac{v_{\rm ej}}{10^3\,\kms}\right)^3\left(\frac{t_{\rm pl}}{100\,\rm d}\right)^4\left(\frac{L_{\rm pl}}{10^{41}{\,\rm erg\,s^{-1}}}\right)^{-1}\ ,
    \label{eq:Mej_popov}\\
R_0&\simeq140\,\Rsun\left(\frac{v_{\rm ej}}{10^3\,\kms}\right)^{-4}\left(\frac{t_{\rm pl}}{100\,\rm d}\right)^{-2}\left(\frac{L_{\rm pl}}{10^{41}{\,\rm erg\,s^{-1}}}\right)^{2}\ .
    \label{eq:R_popov}
\end{align}
Unfortunately, these estimates depend sensitively on the ejecta speed $v_{\rm ej}$, which is challenging to measure and comes with large uncertainties.  Furthermore, the observed precursor duration $t_{\rm pre}$ provides only a lower limit on the total duration ($t_{\rm pl}$) which enters these formulae if the SN prematurely terminates the precursor emission. 

Table~\ref{table:erpt} provides the ejecta mass, kinetic energy ($E_{\rm kin}=M_{\rm ej}v_{\rm obs}^2/2$), and progenitor radius for our precursor sample (Table \ref{table:obs}) as obtained from the inverted Popov formulae (Eqs.~\ref{eq:Mej_popov}, \ref{eq:R_popov}).  Again, the observed precursor duration is only a lower limit on the total plateau duration and hence we obtain only a lower (upper) limit on the ejecta mass (progenitor radius).  All values are calculated assuming lowest spectroscopically observed velocity range.  However, we again caution that our estimates are highly dependent on the adopted ejecta speed; a factor of two difference in the velocity changes the results by an order of magnitude or more.

\begin{table}
\begin{center}
\caption{CSM Properties implied by Eruption Model}
\label{table:erpt}
\begin{tabular}{lccc}
\hline
Event&$M_{\rm ej}^{(a)}$&$E_{\rm kin}^{(a)}$&$R_0^{(b)}$\\
&[$\Msun$]&[$10^{48}$erg]&[$10^{3}\Rsun$]\\
\hline
SN 2009ip&$0.036-0.19$&$0.23-3.8$&$2.4-23$\\
SN 2010mc&$0.039-1.0$&$0.39-94$&$0.046-3.7$\\
SN 2015bh&$1.3-5.9$&$4.6-59$&$0.13-1.0$\\
SN 2016bdu&$0.32$&$0.51$&$8.4$\\
SN 2020tlf&$0.025-1.6$&$0.00062-0.63$&$0.42-110$\\
\hline
\multicolumn{4}{l}{{\bf Note.} All quantities calculated by applying the observed}\\
\multicolumn{4}{l}{precursor properties (Table \ref{table:obs}) to Eqs.~\eqref{eq:Mej_popov}, \eqref{eq:R_popov}, assuming}\\
\multicolumn{4}{l}{the lowest spectroscopically observed velocity (which gives}\\
\multicolumn{4}{l}{the most physical ejecta mass; Fig.~\ref{fig:csm}).}\\
\multicolumn{4}{l}{$^{(a)}$ Lower limit, based on the requirement that $t_{\rm pl} > t_{\rm pre}$.}\\
\multicolumn{4}{l}{$^{(b)}$ Upper limit, based on the requirement that $t_{\rm pl} > t_{\rm pre}$.}\\
\end{tabular}
\end{center}
\end{table}

At the time of the SN explosion, the characteristic outer radius and density of the precursor ejecta (hereafter CSM) are thus given, respectively, by
\begin{align}
R_{\rm CSM}&\sim v_{\rm ej}t_{\rm pre}\simeq8.6\times10^{14}{\,\rm cm\,}\left(\frac{v_{\rm ej}}{10^3\,\rm km\,s^{-1}}\right)\left(\frac{t_{\rm pre}}{100\,\rm d}\right)\ ,
    \label{eq:csm radius}\\
\rho_{\rm CSM}&\sim\frac{M_{\rm ej}}{\frac{4\pi }{3}R_{\rm CSM}^3}\gtrsim5.0\times10^{-12}{\,\rm g\,cm^{-3}\,}
    \nonumber\\
&\times\left(\frac{t_{\rm pre}}{100\,\rm d}\right)\left(\frac{L_{\rm pre}}{10^{41}{\,\rm erg\,s^{-1}}}\right)^{-1}\ ,
    \label{eq:csm density erpt}
\end{align}
where the second equality in Eq.~\eqref{eq:csm density erpt} makes use of Eq.~(\ref{eq:Mej_popov}).  The upper limit on $\rho_{\rm CSM}$ results because $t_{\rm pl}>t_{\rm pre}$ and is notably independent of the (poorly constrained) ejecta velocity.

\begin{figure*}
\begin{center}
\includegraphics[width=165mm, angle=0,bb=0 0 515 239]{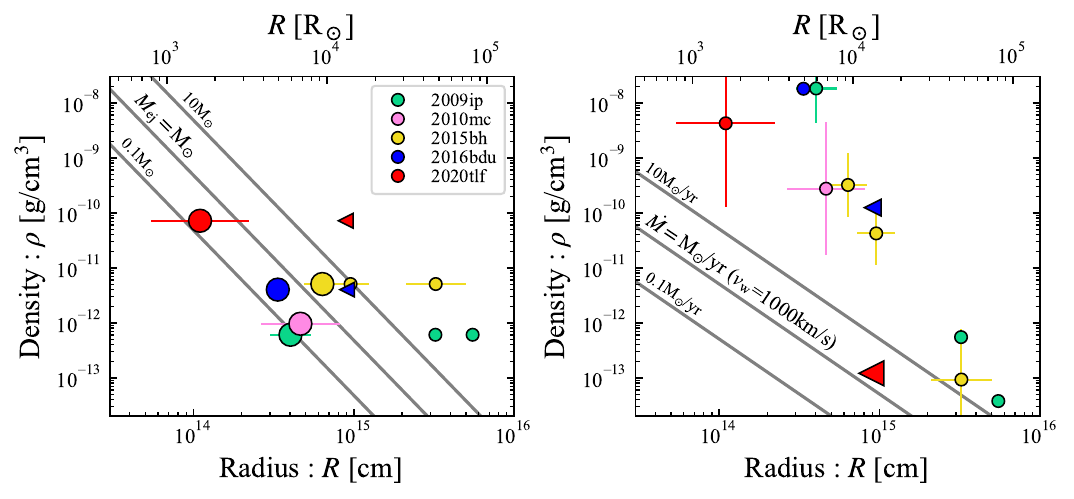}
\caption{CSM density profile resulting from eruption (left) and wind (right) mass-loss models. The gray lines in left (right) panel show corresponding  ejecta mass (mass-loss rate) of $M_{\rm ej}(\dot{M})=0.1$, 1, and 10 $\Msun$ ($\Msun\,\rm yr^{-1}$). For the same event, different points are calculated by different velocity obtained by observations (see Table~\ref{table:obs}).}
\label{fig:csm}
\end{center}
\end{figure*}

The left panel of Fig.~\ref{fig:csm} shows the estimated CSM density $\rho_{\rm CSM}$ and radius $R_{\rm CSM}$ in the eruption scenario for the observed precursors (Fig.~\ref{fig:obs}, Table~\ref{table:obs}).  As discussed in Sec.~\ref{sec:eruption_results} and shown in Fig.~\ref{fig:lc_eruption}, the required CSM masses are typically in the range $\sim 0.1-1\,\Msun$.  Since the light curve properties depend only weakly on the ejecta velocity profile $M_{\rm ej} \propto v^{-\beta}$ (Eq.~\ref{eq:mass_profile}), the slope of the resulting CSM radial density profile $\rho \propto r^{-\beta-3}$ (at $r > R_{\rm CSM}$ or equivalently $v>v_{\rm ej}$) is not well-constrained by the precursor observations.  However, as long as $\beta \gtrsim 0$, the CSM mass will be concentrated in the $\sim v_{\rm ej}$ shell on the radial scale $\sim R_{\rm CSM}$.

Several implications follow from the implied CSM properties.
The inferred CSM is sufficiently massive and compact as to be optically thick ($\rho\gtrsim10^{-14}{\,\rm g\,cm^{-3}\,}(R_{\rm CSM}/10^{14}{\,\rm cm})^{-1}$), thus implying that (1) the SN shock breakout will occur from the precursor-generated CSM shell instead of the original stellar surface; (2) flash spectroscopy cannot readily probe the mean velocity of the precursor ejecta (though it may probe its high-velocity tail).  

Shock heating of the CSM shell by the supernova ejecta will dissipate the kinetic energy of the latter and boost the luminosity of the supernova relative to the CSM-free case.  The luminosity of this shock interaction can be crudely estimated as\footnote{This scaling differs from that for interacting SN with wind-like CSM because of the steeper density profile. } 
\begin{align}
& L_{\rm sh}\sim \frac{\frac{1}{2}\varepsilon M_{\rm ej}v_{\rm SN}^2}{t_{\rm diff}}\frac{R_{\rm CSM}}{v_{\rm SN}t_{\rm diff}} \nonumber \\
&\simeq6\times10^{42}{\,\rm erg\,s^{-1}}\left(\frac{\varepsilon}{0.1}\right)\left(\frac{R_{\rm CSM}}{10^{14}\,\rm cm}\right)\left(\frac{v_{\rm SN}}{10^4\,\rm km\,s^{-1}}\right)^2\ ,
\end{align}
where $t_{\rm diff}=\sqrt{\frac{\kappa M_{\rm ej}}{4\pi cv_{\rm SN}}}$ is the characteristic photon diffusion time through the shocked CSM, and $\varepsilon$ is the radiative efficiency. Interestingly, the luminosity depends only on the CSM radius.  For CSM radii $R_{\rm CSM}\simeq3\times10^{14}\,\rm cm$ implied by the precursor emission (Fig.~\ref{fig:dist_erpt}) the predicted luminosities $L_{\rm sh} \sim10^{43}\,\rm erg\,s^{-1}$ broadly agree with those of the SNe following the precursor events (Fig.~\ref{fig:obs}). However, for SN 2020tlf with $R_{\rm CSM}\sim10^{14}\,\rm cm$, the shock-boosted SN luminosity would be too low and another CSM component, for example optically-thick CSM extending to large radii $\sim 10^{15}\,\rm cm$, is required to power the early SN light curve via CSM interaction.

Though constraints on the CSM properties from the literature of our precursor sample are limited, we comment briefly on those available.  For SN 2009ip, \citet{Margutti+14,Moriya2015} interpret the SN emission as arising due to shock breakout emission from a dense shell produced by the precursor outburst; they estimate a precursor mass of $\sim0.1\,\Msun$ and radius $\sim10^{15}\,\rm cm$ at explosion, by equating the SN rise-time to the photon diffusion time through the dense shell.  Our result is essentially consistent with theirs for this event, for an assumed ejecta velocity equal to the lowest measured spectroscopically, $v_{\rm ej}\simeq 800-1400\,\kms$. For SN 2020tlf, \cite{Chugai&Utrobin2022} model the precursor emission as arising from an ejecta mass of $M_{\rm ej}\sim0.1\,\Msun$, also consistent with our result. They find that subsequent SN emission is not affected by this compact CSM, but that an additional wind-like CSM extending up to $\sim10^{15}\,\rm cm$ is required (we return to this event within the context of the wind-scenario in the next section).

\subsection{Wind scenario}
Detecting the rising phase of the light curve in the wind scenario is typically challenging (see Fig.~\ref{fig:lc_wind}).  We therefore assume that the emission has already reached its stationary state by the time of the observations.  Inverting Eq.~\eqref{eq:LlowMdot}, we derive the required wind mass-loss rate in terms of observed precursor properties:
\begin{align}
\dot{M}&\simeq1.5{\,\smyr\,}\left(\frac{M_\star}{10\,\Msun}\right)^{-2}
    \nonumber\\
&\times\left(\frac{v_{\rm w}}{10^3{\,\rm km\,s^{-1}}}\right)^{-2}\left(\frac{L_{\rm pre}}{10^{40}{\,\rm erg\,s^{-1}}}\right)^{3}\ .
\end{align}
Insofar as the outer radius of the wind ejecta at the time of the SN explosion is given by $R_{\rm CSM}\sim v_{\rm w}t_{\rm active}\gtrsim v_{\rm w}t_{\rm pre}$, Eq.~\eqref{eq:csm radius} with $v_{\rm ej}$ replaced by $v_{\rm w}$ gives a lower limit on $R_{\rm CSM}$ while the rise-time is typically shorter than $t_{\rm pre}$ for relevant wind velocities. This in turn implies an upper limit on the CSM density at $R_{\rm CSM}$,
\begin{align}
\rho_{\rm CSM}&=\frac{\dot{M}}{4\pi R_{\rm CSM}^2v_{\rm w}}\\
&\lesssim1.0\times10^{-13}{\,\rm g\,cm^{-3}\,}\left(\frac{M_\star}{10\,\Msun}\right)^{-2}
\nonumber\\
&\times\left(\frac{v_{\rm w}}{10^3{\,\rm km\,s^{-1}}}\right)^{-5}\left(\frac{L_{\rm pre}}{10^{40}{\,\rm erg\,s^{-1}}}\right)^{3}\left(\frac{t_{\rm pre}}{100\,\rm d}\right)^{-2}\ .
	\label{eq:csm density wind}
\end{align}
which depends sensitively on the wind velocity. Assuming the wind mass-loss rate is roughly constant prior to the SN, the CSM density at $r < R_{\rm CSM}$ will follow a wind-like $\rho\propto r^{-2}$ profile.

The right panel of Fig.~\ref{fig:csm} shows the estimated CSM density of observed precursors in the wind mass-loss scenario.  As discussed in Sec.~\ref{sec:wind_results}, the required CSM density, or equivalently mass-loss rate $\dot{M}$, is enormous, exceeding that allowed by the total mass budget of the star in most events. 

Only in SN 2020tlf can the observed precursor be reasonably accommodated, for the maximally-allowed velocity $v_{\rm w}\sim v_{\rm engulf}\simeq10^3\,\kms$ (and we note that $\dot{M} \simeq const$ mass-loss may indeed be suggested for this event by the flat shape of the precursor light curve).  The implied radial extent of the CSM $\sim10^{15}\,\rm cm$ in the wind scenario is consistent with an upper limit based on X-ray non-detections \citep{JacobsonGalan+2022} and light-curve modeling \citep{Chugai&Utrobin2022} (though we note that the density required in our scenario is $3-10$ times greater than found by these authors).  Given uncertainties on parameters, such as the progenitor mass and wind velocity, and potential for internal-shock-heating (Sec.~\ref{sec:shocks}), this level of discrepancy may nevertheless be tolerable.  We note that the CSM radius obtained for SN 2020tlf in the eruption scenario (Table \ref{table:erpt}) is also consistent with the modeling of \citet{Chugai&Utrobin2022}.

As in the eruption case, the inferred CSM density (wind mass-loss rate) should not be so large that it overproduces the SN luminosity via CSM shock interaction.  For freely expanding SN ejecta into a wind-like density profile, the SN luminosity can be estimated as (e.g., \citealt{Chevalier&Irwin11}, but here ignoring deceleration of the shocked CSM shell):
\begin{align} 
L_{\rm sh}&\sim \frac{\frac{1}{2}\varepsilon M_{\rm swpt}v_{\rm SN}^2}{t_{\rm diff}}\\
&\simeq3\times10^{43}{\,\rm erg\,s^{-1}}\left(\frac{\varepsilon}{0.1}\right)\left(\frac{\dot{M}}{\Msun\,\rm yr^{-1}}\right)
    \nonumber\\
&\times\left(\frac{v_{\rm SN}}{10^4\,\rm km\,s^{-1}}\right)^3\left(\frac{v_{\rm w}}{10^3{\,\rm km\,s^{-1}}}\right)^{-1}\ ,
\end{align}
where $M_{\rm swpt}\sim \rho R_{\rm diff}^3$ is the CSM wind mass swept-up by SN ejecta on the wind diffusion timescale $t_{\rm diff}=\frac{\kappa\dot{M}}{4\pi cv_{\rm w}}$ (Eq.~\ref{eq:t_diff}), and we assumed the radius $R_{\rm diff} \sim v_{\rm SN}t_{\rm diff}\sim6\times10^{14}{\,\rm cm\,}(\dot{M}/\smyr)(v_{\rm SN}/10^4{\,\rm km\,s^{-1}})(v_{\rm w}/10^3\,\rm km\,s^{-1})^{-1}$ is smaller than the outer edge of the CSM, $R_{\rm CSM}$.  

The estimate for $L_{\rm sh}$ in the case of SN 2020tlf is roughly consistent with its SN luminosity, while we find $L_{\rm sh} \gg L_{\rm SN}$ given the much higher values $\dot{M}$ required to explain other precursors.  This again favors the eruption scenario over the steady-wind scenario as the origin of most SN precursors, at least neglecting shock-heating within the wind (Sec.~\ref{sec:shocks}).

\section{Summary}\label{sec:summary}

We have modeled the precursor optical emission detected from a growing sample of core collapse SNe using an extension of the semi-analytical light curve model described in \cite{Matsumoto&Metzger2022}.  The observed precursors can be regarded as extreme cases of the mass-loss events increasingly inferred to occur just before the terminal collapse, whose direct emission can provide clues to the nature of the mass-loss mechanisms at work and more generally about the final stages of massive star evolution.  We develop light curve models in the context of two scenarios  (`eruption' and `wind') for the nature of the pre-SN mass-loss phase.  

In the eruption model, energy is deposited near the base of stellar envelope on a timescale shorter than the dynamical time.  While our model is indifferent to the mechanism responsible for this energy injection, it could in principle be caused by a violent outburst associated with late stages of nuclear shell burning.  This sudden energy deposition creates a shock wave, which propagates outwards towards the surface of the star, ejecting a portion of its envelope.  For the eruption scenario, our findings are summarized as follows:
\begin{itemize}
\item The eruption light curve is characterized by a recombination-driven plateau similar to that of Type IIP SNe (Fig.~\ref{fig:lc_eruption}).  The luminosity and duration of the plateau are reasonably estimated by the so-called ``Popov formulae'' \citep{Popov1993} which relates these observables to the ejecta mass and progenitor radius (Eqs.~\ref{eq:L_popov}, \ref{eq:t_popov}).  However, because the entirety of precursor light curve is generally not observable (due to interruption by the SN explosion), we can obtain only lower limits on the ejecta mass $M_{\rm ej}$ and kinetic energy $E_{\rm kin}$ for individual events.  Adopting an ejecta velocity $v_{\rm obs}\sim100-1000\,\rm km\,s^{-1}$ consistent with those inferred from spectroscopy, the observed precursor luminosities are reproduced for $M_{\rm ej}\gtrsim0.1-1\,\Msun$ and $E_{\rm kin}\gtrsim10^{48}-10^{50}\,\rm erg$ (Table~\ref{table:erpt}).  We caution that the latter estimate depends sensitively on the ejecta velocity, which is not available in all events and could in principle be confused with CSM from other, earlier phases of mass-loss. 

\item A wide range of precursor luminosities can be produced in this model, $L_{\rm pre}\sim10^{39}-10^{42}\,\rm erg\,s^{-1}$ (Fig.~\ref{fig:dist_erpt}, see also \citealt{Dessart+2010}). The luminosity mainly depends on the progenitor radius and ejecta velocity. Bright precursors of $L_{\rm pre}\gtrsim10^{41}\,\rm erg\,s^{-1}$ require large progenitor radii $R_\star\gtrsim 10^2\,\Rsun$ such as RSGs.  For these progenitors, super-Eddington luminosities $L_{\rm pre}\gtrsim L_{\rm Edd}\sim10^{39}\,\rm erg\,s^{-1}\,(M_\star/10\,\Msun)$ are achieved only for the ejecta speeds larger than the escape speed of the stellar surface, supporting the source of energy deposition occurring deep inside the stellar envelope.
\end{itemize}

The second scenario we considered is one of steady wind-like mass-loss, which results from a continuous energy deposition at the base of stellar envelope (mainly motivated by the wave-heating scenario; e.g., \citealt{Quataert&Shiode2012,Fuller2017}). For the wind scenario, our findings are summarized as follows:
\begin{itemize}
\item The light curve undergoes a gradual rise, before settling into a stationary state. The rise-time is essentially given by the diffusion timescale of the wind, and the steady-state luminosity as $L\sim L_{\rm Edd}(\frac{1}{2}\dot{M}v_{\rm w}^2/L_{\rm Edd})^{1/3}$ (Eq.~\ref{eq:LlowMdot}, \ref{eq:LhighMdot}; see also \citealt{Quataert+2016,Shen+2016}).  For typical values of the observed precursor ejecta velocity, very large mass-loss rates $\dot{M}\gtrsim10\,\Msun\,\rm yr^{-1}$ are required to achieve the observed precursor luminosities $L_{\rm pre}\sim10^{41}\,\rm erg\,s^{-1}$ (Fig.~\ref{fig:dist_wind}).  Such large mass-loss rates would also increase the light curve rise-time which (in conjunction with the large required $\dot{M}$) would exceed the mass budget of the star.  This constraint can be formalized by defining a theoretical maximum precursor luminosity for a given wind velocity (Fig.~\ref{fig:Lmax_wind}; Appendix \ref{sec:appendix}). 

Among the five well-observed precursors in our sample, only SN 2020tlf with its relatively low luminosity, can be explained by the wind model, for a mass-loss rate of $\dot{M}\simeq1\,\Msun\rm\,yr^{-1}$ and velocity of $v_{\rm w}\simeq10^3\,\rm km\,s^{-1}$. The corresponding sonic radius is again deep inside the star, $R_{\rm s}\sim10\,\Rsun\ll10^2-10^3\,\Rsun$, requiring energy deposition at the base of stellar envelope.
\item The luminosity of the wind can be boosted due to internal shocks, for instance due to a time-variable wind speed, by a modest factor $\lesssim 2$. However, the relatively smooth observed shape of the precursor light curves may disfavor shock interaction. 

While we do not study them in detail, external shocks between the precursor ejecta and pre-existing CSM may also take place and contribute to the luminosity \citep[see also][]{Quataert+2016,Strotjohann+21,JacobsonGalan+2022}. 
The shock luminosity can be crudely estimated by energy and momentum conservation. Neglecting any losses of efficiency due to deceleration of wind, the shock-dissipated energy is given by $E_{\rm sh}\sim\frac{1}{2}\left(\frac{M_{\rm w}M_{\rm pCSM}}{M_{\rm w}+M_{\rm pCSM}}\right)v_{\rm w}^2$, where $M_{\rm w}=\dot{M}t_{\rm pre}$ is the wind mass ejected during the precursor and $M_{\rm pCSM}$ is the pre-existing CSM mass swept up by the wind. For pre-existing CSM with a wind-like profile, the swept-up mass is given by $M_{\rm pCSM}=\int^{v_{\rm w}t_{\rm pre}}4\pi r^2 \rho_{\rm pCSM}dr=\dot{M}_{\rm pCSM}v_{\rm w}t_{\rm pre}/v_{\rm pCSM}$, where $\dot{M}_{\rm pCSM}$ and $v_{\rm pCSM}$ are the mass-loss rate and velocity of the pre-existing CSM, respectively. Combining results, the radiated shock luminosity is given by 
\begin{align}
L_{\rm sh}\sim\frac{E_{\rm sh}}{t_{\rm pre}}\sim\left(\frac{1}{1+r_{\rho}}\right)\dot{E}_{\rm w}\ ,
	\label{eq:Lsh}
\end{align}
where $r_\rho=\rho_{\rm w}/\rho_{\rm pCSM}$ is independent of $r$ for wind-like profiles. While $r_{\rho}>1$ is required to neglect deceleration of the wind, and hence $L_{\rm sh} \lesssim \dot{E}_{\rm w}$, the stronger dependence of the radiated luminosity on $\dot{E}_{\rm w}$ in Eq.~\eqref{eq:Lsh} than Eq.~\eqref{eq:LlowMdot} suggests that shock interaction can still dominate over the intrinsic wind luminosity.  Future work is needed to explore the external shock scenario in greater detail.

\item Absent external shock interaction, the fact that observed SN precursors violated the maximal luminosity in the wind scenario may favors eruptive events (e.g., driven by unstable nuclear shell burning) over a more continuous heating source (e.g., due to wave heating).  
\end{itemize}

In both eruption and wind models, the precursor ejecta forms an optically-thick compact CSM shell surrounding the progenitor star that will be present at the time of collapse and could boost the luminosity of the supernova light curve substantially through shock interaction. Our main finding is:
\begin{itemize}
\item The radial extent of the CSM is typically so small $R_{\rm CSM}\lesssim10^{15}\,\rm cm$ (see Eq.~\ref{eq:csm radius}) that it is engulfed by SN ejecta less than ten days after SN explosion. Therefore, flash spectroscopy carried out a few days after SN may miss the signature of the precursor ejecta and probe only the CSM released before the precursor mass-loss episode \citep[see also][]{Strotjohann+21}. In the eruption model, the density profile of the CSM is steeper than the steady wind profile ($\rho\propto r^{-2}$), forming a dense CSM core.
\end{itemize}

\acknowledgements
We thank Avishay Gal-Yam and Eliot Quataert for helpful comments on an early draft of the text.  
This work is supported in part by JSPS Overseas Research Fellowships (T.M.).  B.D.M. acknowledges support from the National Science Foundation (grant number AST-2009255).

\appendix

\section{Analytic Estimates of Precursor Emission in the Wind Mass-Loss Scenario}
\label{sec:appendix}

The rise-time of the precursor light curve (equivalently, the duration of the single shell light curve; Fig.~\ref{fig:lc_wind}) is determined by the diffusion timescale through the shell, $t_{\rm d}$ (Eq. \ref{eq:tdwind}).  Radiation thus escapes the wind ejecta from the characteristic radius
\begin{align}
R_{\rm d} \simeq v_{\rm w}t_{\rm d}=\frac{\kappa\dot{M}}{4\pi c}\simeq5.4\times10^{13}{\,\rm cm\,}\biggl(\frac{\dot{M}}{\smyr}\biggl)\ .
    \label{eq:r_diff}
\end{align}
In this appendix, we normalize $\kappa$ to the fully-ionized electron scattering opacity, $\kappa_{\rm es}\simeq0.32\,\rm cm^2\,g^{-1}$.  Because photons are trapped, the wind material expands adiabatically between the sonic and diffusion radii ($R_{\rm s} < r < R_{\rm d}$), such that $T\propto \rho^{1/3} \propto r^{-2/3}$ under radiation-pressure-dominated condition and $\rho \propto r^{-2}$ for a steady wind.  The thermal and kinetic luminosities of the wind are comparable at $R_{\rm s}$; thus, the photon (advection) luminosity $L \simeq 4\pi r^{2}v_{\rm w} aT^4\propto T^{4}r^{2} \propto r^{-2/3}$ evolves as
\begin{align}
L&\simeq\dot{E}_{\rm w}\biggl(\frac{r}{R_{\rm s}}\biggl)^{-2/3}\ ,
	\label{eq:wind luminosity}
\end{align} 
up to the diffusion becomes important. Here $\dot{E}_{\rm w}$ is the wind kinetic power (Eq.~\ref{eq:wind_luminosity}).

There are two cases to consider, depending on whether the wind material remains fully ionized or has begun to recombine below $\sim R_{\rm d}$.
When $\dot{M}$ is sufficiently low (the precise threshold, $\dot{M}_{\rm ion}$, to be defined below), the ejecta is still fully ionized ($T \gtrsim T_{\rm ion} \simeq 10^{4}$ K) when photons start to diffuse our of the wind. Inserting Eq.~\eqref{eq:r_diff} into Eq.~\eqref{eq:wind luminosity}, we find
\begin{align}
L=L_{\rm Edd}\biggl(\frac{\dot{E}_{\rm w}}{L_{\rm Edd}}\biggl)^{1/3}
\simeq3.2\times10^{39}{\,\lumi\,}M_{\star,1}^{2/3}\biggl(\frac{\dot{M}}{\smyr}\biggl)^{1/3}\biggl(\frac{v_{\rm w}}{200\,\kms}\biggl)^{2/3}\ ,  \dot{M} < \dot{M}_{\rm ion},
\label{eq:LlowMdot}
\end{align}
where $M_{\star,1} = M_{\star}/(10M_{\odot}).$  This expression agrees with previous works \citep[e.g.,][]{Quataert+2016,Shen+2016}.

On the other hand, when $\dot{M}$ is sufficiently large, hydrogen recombination begins below $R_{\rm d}$, which triggers photon diffusion by reducing the opacity $\kappa \ll \kappa_{\rm es}$ and effectively shrinking the diffusion radius. Recombination occurs at the temperature $T = T_{\rm ion} = 10^{4}T_{\rm ion,4}$ K, as is achieved at the radius
\begin{align}
R_{\rm ion}=R_{\rm s}\biggl(\frac{T_{\rm s}}{T_{\rm ion}}\biggl)^{3/2}=\biggl(\frac{\dot{M}^3(2GM_\star)^2}{(8\pi a T_{\rm ion}^4)^3v_{\rm w}}\biggl)^{1/8}
\simeq2.5\times10^{14}{\,\rm cm\,}M_{\star,1}^{1/4}T_{\rm ion,4}^{-3/2}\biggl(\frac{\dot{M}}{\smyr}\biggl)^{3/8}\biggl(\frac{v_{\rm w}}{200\,\kms}\biggl)^{-1/8}\ ,
\end{align}
on the timescale
\begin{align}
t_{\rm ion} =\frac{R_{\rm ion}}{v_{\rm w}}\simeq140{\,\rm d\,}M_{\star,1}^{1/4}T_{\rm ion,4}^{-3/2}\biggl(\frac{\dot{M}}{\smyr}\biggl)^{3/8}\biggl(\frac{v_{\rm w}}{200\,\kms}\biggl)^{-9/8}\ ,
    \label{eq:tion}
\end{align}
where
\begin{align}
T_{\rm s}=\biggl(\frac{\dot{M}v_{\rm w}}{8\pi a R_{\rm s}^2}\biggl)^{1/4}=\biggl(\frac{\dot{M}v_{\rm w}^5}{8\pi a (2GM_\star)^2}\biggl)^{1/4} \nonumber 
\simeq1.1\times10^{5}{\,\rm K\,}M_{\star,1}^{-1/2}\biggl(\frac{\dot{M}}{\smyr}\biggl)^{1/4}\biggl(\frac{v_{\rm w}}{200\,\kms}\biggl)^{5/4}\ 
\end{align}
is the wind temperature at the sonic radius.  The wind luminosity in the high-$\dot{M}$ case (Eq.~\ref{eq:wind luminosity}) is then given by
\begin{align}
L&= \dot{E}_{\rm w}\left(\frac{R_{\rm ion}}{R_{\rm s}}\right)^{-2/3} = L_{\rm Edd}\biggl(\frac{\dot{E}_{\rm w}}{L_{\rm Edd}}\biggl)^{1/3}\biggl(\frac{\dot{M}}{\dot{M}_{\rm ion}}\biggl)^{5/12}=\biggl(\frac{\pi  a T_{\rm ion}^4(2GM_\star)^2\dot{M}^{3}v_{\rm w}^{3}}{2}\biggl)^{1/4}\ , \nonumber \\
&\simeq1.2\times10^{39}{\,\rm erg\,}M_{\star,1}^{1/2}T_{\rm ion,4}\biggl(\frac{\dot{M}}{\smyr}\biggl)^{3/4}\biggl(\frac{v_{\rm w}}{200\,\kms}\biggl)^{3/4}\,\,\, ,\dot{M} > \dot{M}_{\rm ion} .
\label{eq:LhighMdot}
\end{align}
The critical mass-loss rate above which recombination defines the transient luminosity can be estimated by equating Eqs.~(\ref{eq:LlowMdot}) and (\ref{eq:LhighMdot}),
\begin{align}
\dot{M}_{\rm ion}=\biggl(\frac{2^7\pi^5(2GM_\star)^2c^8}{a^3\kappa^8T_{\rm ion}^{12}v_{\rm w}}\biggl)^{1/5}
\simeq11{\,\smyr\,}M_{\star,1}^{2/5}T_{\rm ion,4}^{-12/5}\biggl(\frac{v_{\rm w}}{200\kms}\biggl)^{-1/5}\ .
\label{eq:Mdotcri}
\end{align}

For precursor emission which exhibits a steady-state plateau-like light curve, the mass-loss must be active for a timescale $\gtrsim t_{\rm d}$ ($t_{\rm ion}$). For a given allowed ejecta mass $M_{\rm ej}$, this implies a maximum on the wind mass-loss rate can be obtained. For the low-$\dot{M}$ case, demanding $t_{\rm d}\dot{M}<M_{\rm ej}$ gives
\begin{align}
\dot{M}_{\rm max}=\left(\frac{4\pi cv_{\rm w}M_{\rm ej}}{\kappa}\right)^{1/2}
\simeq11{\,\smyr\,}M_{\star,1}^{1/2}\left(\frac{M_{\rm ej}}{M_\star}\right)^{1/2}\left(\frac{v_{\rm w}}{200\kms}\right)^{1/2}\ ,
    \label{eq:Mdot_max1}
\end{align}
and corresponding maximal steady-state luminosity is given by
\begin{align}
L_{\rm max}=7.0\times10^{39}{\lumi}M_{\star,1}^{5/6}\left(\frac{M_{\rm ej}}{M_\star}\right)^{1/6}\left(\frac{v_{\rm w}}{200\kms}\right)^{5/6}\ .
\label{eq:Lmax1}
\end{align}
On the other hand, for the low $\dot{M}$ case, the maximal mass-loss rate and luminosity are given by
\begin{align}
\dot{M}_{\rm max}&=\left(\frac{(8\pi a T_{\rm ion}^4)^3v_{\rm w}^9M_{\rm ej}^8}{(2GM_\star)^2}\right)^{1/11}
\simeq11{\,\smyr\,}M_{\star,1}^{6/11}T_{\rm ion,4}^{12/11}\left(\frac{M_{\rm ej}}{M_\star}\right)^{8/11}\left(\frac{v_{\rm w}}{200\kms}\right)^{9/11}\ ,\\
L_{\rm max}&=6.8\times10^{39}{\lumi}M_{\star,1}^{10/11}T_{\rm ion,4}^{20/11}\left(\frac{M_{\rm ej}}{M_\star}\right)^{6/11}\left(\frac{v_{\rm w}}{200\kms}\right)^{15/11}\ .
\label{eq:Lmax2}
\end{align}

\bibliographystyle{aasjournal}
\bibliography{reference_matsumoto,refs_BDM}

\end{document}